\documentclass[jcp,aps,preprint]{revtex4}
\usepackage{amsmath}
\usepackage{amstext,bm}
\usepackage{epsfig}
\usepackage{xcolor}
\usepackage{multirow}
\usepackage{rotating}
\definecolor{dgreen}{rgb}{0,.5,0}

\begin{document}
\def\etal{\textit{et al.}}
\def\mup{$\mathrm{\mu}$}
\def\muopt{$\mathrm{\mu_{\rm opt}}$}
\def\imp{$\mathrm{i}$}

\title{Multi-configuration time-dependent density-functional theory \\
based on range separation}
\author{Emmanuel Fromager$^{a}$, Stefan Knecht$^b$ and Hans J{\o}rgen Aa.~Jensen$^b$
\\
}


\affiliation{\it 
\\
$^a$Laboratoire de Chimie Quantique,\\
Institut de Chimie, CNRS / Universit\'{e} de Strasbourg,\\
4 rue Blaise Pascal, 67000 Strasbourg, France.\\
\\
$^b$Department of Physics, Chemistry and Pharmacy, University of Southern Denmark,\\
Campusvej 55, DK-5230 Odense M, Denmark.
}


\begin{abstract}

Multi-configuration range-separated density-functional theory is
extended to the time-dependent regime. An exact variational formulation
is derived. The approximation, which consists in combining a long-range {\it
Multi-Configuration-Self-Consistent Field} (MCSCF) treatment with an
adiabatic 
short-range density-functional (DFT) description, is then considered. 
The resulting time-dependent
multi-configuration short-range DFT (TD-MC-srDFT) model is 
applied to the calculation of singlet excitation energies in H$_2$, Be and
ferrocene, considering both short-range local density (srLDA) and generalized gradient (srGGA)
approximations. 
In contrast to regular TD-DFT, TD-MC-srDFT can describe
double excitations. 
As expected, when modeling long-range interactions
with the MCSCF model instead of the
adiabatic Buijse-Baerends density-matrix functional as recently
proposed by Pernal [K.~Pernal, J. Chem. Phys. {\bf 136}, 184105 (2012)], the description of  
both the $1^1D$ doubly-excited state in Be 
and the $1^1\Sigma^+_u$ state in the stretched H$_2$ molecule are improved,
although the latter is still significantly underestimated. 
Exploratory TD-MC-srDFT/GGA calculations 
 for ferrocene yield in general excitation energies at least as good as TD-DFT/CAM-B3LYP,
 and superior to wave-function (TD-MCSCF, 
 symmetry adapted cluster-configuration interaction) and 
 TD-DFT results based on LDA, GGA, and hybrid functionals.  

\end{abstract}
\maketitle

\clearpage

\section{Introduction}

Time-dependent density-functional theory
(TD-DFT)~\cite{PRL84_RGth,Casida_eqs} is used routinely nowadays for
computing electronic excitation energies and transition properties of molecules and solids. 
Even though low-lying excitation energies often can be obtained sufficiently accurate at a relatively
low computational cost, adiabatic TD-DFT using pure functionals usually fails in
describing, for example, charge transfers and double excitations.
While charge transfers often can  be modeled adequately with range-separated hybrid
functionals~\cite{savinbook,scuseriacalib2,hiraonewmu,camb3lyp_diagnostic,dft-Rohrdanz-JCP2009-130-054112},
where the long-/short-range decomposition of the electron-electron repulsion is used only
for the exchange energy, the double excitations remain problematic 
for all standard exchange-correlation functionals 
as long as the adiabatic approximation is
used for the time-dependent exchange and correlation density-functional~\cite{Elliott_tddft_2ble_XE,maitra_tddft_2ble_XE,neugebauer_tddft_2ble_XE,burke:150901}.
This statement holds for hybrid functionals even though non-adiabatic
effects can be taken into account through the exact exchange
functional, due to
the fact that the Kohn-Sham (KS) orbitals are non-local (in
time) functionals of the density~\cite{PRL02_Burke_memory_tddft,PhysRevLett.89.096402,PhysRevLett.108.146401}. 
The accurate description of charge-transfer states in molecular complexes is of 
particular interest because of their prominent role in for example organic electronics \cite{organic_el_primer_05}\ 
and dye-sensitized solar cell applications \cite{graetzel_dssc_rev_2005}.
To this end, more sophisticated functionals such as LRC-BOP
\cite{hiraomu, hiraonewmu}, CAM-B3LYP \cite{dft-Yanai-CPL2004a}, as well as LC-$\omega$-PBE \cite{scuseriacalib1}\ and 
other types of range-separated hybrids have been developed
\cite{tuned_rsh_rev2010,peach:044118,doi:10.1021/ja8087482}.

In order to further improve standard TD-DFT, in particular for the description
of double excitations, Pernal~\cite{JCP12_Pernal_tddmft-srdft} 
recently proposed to combine it with time-dependent
density-matrix functional theory (TD-DMFT) by means of range
separation. While describing the long-range part of the
electron-electron interaction with the Buijse-Baerends (BB) density-matrix
functional, the short-range interaction is treated within the local
density approximation (srLDA), both within the adiabatic approximation.
The author has tested on H$_2$ and
Be, and she found that the combined method performs much better than
standard TD-LDA and TD-DMFT-BB for many excitations, provided the range separation parameter
is properly chosen. 
However, large errors were obtained for
the $1^1D$ double excitation in Be as well as for the $1^1\Sigma^+_u$
state in the stretched H$_2$ molecule. As an alternative to the
long-range adiabatic TD-DMFT-BB treatment, 
we propose in this work to describe the long-range interaction at the time-dependent 
{\it Multi-Configuration Self-Consistent Field} (TD-MCSCF) level, 
extending thus the MC-srDFT~\cite{JCPunivmu} model to the time-dependent regime.

The paper is organized as follows: after a short introduction to
time-independent range-separated DFT 
(Sec.~\ref{sec_srdft}), an exact  
and variational time-dependent formulation is presented 
in Sec.~\ref{vp_section}. The approximate linear response TD-MC-srDFT scheme, which
consists in describing the long-range interaction at the TD-MCSCF level and
the short-range interaction within adiabatic TD-DFT, is then introduced
and discussed
in Secs.~\ref{subsec_MCtddft} and \ref{sec_interp_poles}. 
The new scheme is applied to two widely varied types of atomic and molecular systems. 
We first study the paradigm systems H$_2$ and Be to illustrate fundamental
benefits of the TD-MC-srDFT approach by comparing to other suggestions for how to go beyond Kohn-Sham DFT. 
We further demonstrate the performance of the new TD-MC-srDFT scheme in an investigation of 
the low-lying singlet excitations of ferrocene.
Thus, following the computational details
(Sec.~\ref{sec-compdetails}), numerical results obtained for singlet
excitations in H$_2$ (Secs.~\ref{subsec_SigmaU} and
\ref{subsec_SigmaG}), Be (Sec.~\ref{subsec_Be}) are discussed. In Sec.~\ref{subsec_ferrocene} we elaborate on valence- and charge-transfer 
singlet excitations in the transition-metal compound ferrocene before drawing conclusions in Sec.~\ref{conclu}.

\section{Theory}\label{THEORY}

This section deals with the rigorous formulation of variational 
range-separated TD-DFT models where the range separation is used for
both exchange and correlation energies. Introducing first
range-separated DFT in Sec.~\ref{sec_srdft}, its extension to the
time-dependent regime is then presented in Sec.~\ref{vp_section}.
In particular, the derivation of an exact variational formulation is
detailed, in
the light of Vignale's recent
work~\cite{PRA08_Vignale_TDVP,PRA11_Vignale_TDVP_reply_Schirmer} on the Runge and Gross variational
principle in TD-DFT. For practical calculations, an adiabatic variational
formulation is also given. The approximate TD-MC-srDFT model, where the
long-range interaction is described within the TD-MCSCF approach, is
then derived in Sec.~\ref{subsec_MCtddft} using Floquet theory. The
interpretation of the TD-MC-srDFT poles as excitation energies is then discussed in
Sec.~\ref{sec_interp_poles}.\\
The theory in this section is completely general in the sense that
it is valid for any reasonable range-separation of the two-electron repulsion,
however, we will for simplicity describe the theory with the most commonly used
choice for the range-separation:
\begin{eqnarray}
w^{\rm lr,\mu}_{ee}(r_{12})&=&{{\rm erf}(\mu r_{12})}/r_{12}, \nonumber\\
w^{\rm sr,\mu}_{ee}(r_{12})&=&{{\rm erfc}(\mu r_{12})}/r_{12} = 1/r_{12} - w^{\rm lr,\mu}_{ee}(r_{12}) ,
\end{eqnarray}
which depends on the parameter $\mu$.
We note that with this choice one has a continuous class of different range-separations when varying $\mu$ from zero to infinity: 
in the $\mu = 0$ limit the long-range part is zero, $w^{\rm lr,0}_{ee}(r_{12})=0 ,$ 
while in the $\mu =+\infty$ limit the short-range part is zero, $w^{\rm
sr,+\infty}_{ee}(r_{12})=0 .$

\subsection{Range-separated density-functional theory}\label{sec_srdft}

In multi-determinant range-separated DFT, 
which in the remainder of this paper is simply referred to as short-range DFT (srDFT), 
the exact ground-state energy of an electronic system can be expressed as 
\begin{eqnarray}\label{energyminpsi}\begin{array} {l}
{\displaystyle
E = \underset{\Psi}{\rm min}\left\{ \langle \Psi\vert
\hat{T}+\hat{W}^{\rm lr,\mu}_{ee}+\hat{V}_{\rm ne}\vert\Psi\rangle
+E^{\rm sr,\mu}_{\rm Hxc}[n_{\Psi}]\right\},  
}
\end{array}
\end{eqnarray}
where $\hat{T}$ is the kinetic energy operator, 
$\hat{W}^{\rm lr,\mu}_{ee}$ denotes the long-range two-electron interaction,
$\hat{V}_{\rm ne}$ is the nuclear potential operator,
and $E^{\rm sr,\mu}_{\rm Hxc}[n]$ denotes
the $\mu$-dependent short-range Hartree-exchange-correlation (srHxc) 
density-functional which describes the complementary short-range
energy
\cite{TouSav_srlrcorr,erferfgaufunc}. The minimizing wave function
$\Psi^{\mu}$ in Eq.~(\ref{energyminpsi}) fulfills the following
self-consistent equation: 
\begin{eqnarray}\label{timeindepsrdfteq}\begin{array} {l}
\displaystyle
\left(\hat{T}+\hat{W}^{\rm lr,\mu}_{ee}+\hat{V}_{\rm ne}+\hat{V}_{\rm Hxc}^{\rm sr,\mu}[n_{\Psi^{\mu}}]\right) \vert\Psi^{\mu}\rangle=\mathcal{E}^{\mu}\vert\Psi^{\mu}\rangle,\\
\\
\displaystyle
\hat{V}_{\rm Hxc}^{\rm sr,\mu}[n]=\int d\mathbf{r}\;\frac{\delta{E^{\rm
sr,\mu}_{\rm Hxc}}}{\delta n(\mathbf{r})}[n]\;\hat{n}(\mathbf{r}),
\end{array}
\end{eqnarray}
where $\hat{n}(\mathbf{r})$ is the density operator. 
It is readily seen
from Eqs.~(\ref{energyminpsi}) and (\ref{timeindepsrdfteq}) that srDFT
reduces to standard Kohn-Sham (KS) DFT and pure wave function theory (WFT) in the
$\mu=0$ and $\mu\rightarrow+\infty$ limits, respectively. A combined
WFT-srDFT approach is then obtained for intermediate $\mu$ values. 
The approximate Hartree-Fock-srDFT (HF-srDFT) scheme consists in
restricting the minimization in
Eq.~(\ref{energyminpsi}) over single-determinant wave
functions. It is very similar to range-separated hybrid
schemes where the range separation is only used for the
exchange
energy~\cite{scuseriacalib2,hiraonewmu,camb3lyp_diagnostic,dft-Rohrdanz-JCP2009-130-054112}.
The two approaches differ by the fact that HF-srDFT does
not describe long-range correlation effects since, in srDFT, the range separation is also
used for the correlation energy. In the case of
multi-configurational systems, the minimization in
Eq.~(\ref{energyminpsi}) should be performed over 
MCSCF-type wave functions 
instead. This scheme is referred to 
as MC-srDFT~\cite{jesperthesis,JCPunivmu} in the following.
Note that correlated methods such as second-order M\o ller-Plesset
perturbation theory (MP2)~\cite{srdftnancy}, Random-Phase Approximation
(RPA)~\cite{rpa-srdft_toulouse,rpa-srdft_scuseria},
Coupled-Cluster (CC)~\cite{ccsrdft}
and second-order $n$-electron valence state perturbation
theory (NEVPT2)~\cite{nevpt2srdft} have also been applied in this context, using the
non-variational formulation of srDFT which is given in
Eq.~(\ref{timeindepsrdfteq}). 
It is
important to mention that, when describing the long-range interaction
at the post-HF level and the short-range interaction
with a short-range density-functional, there is no risk of double counting electron correlation
effects, which is essential for a combined 
post-HF-DFT approach. In this respect, using a range-separation of the
two-electron repulsion is
very appealing. On the other hand, the performance of post-HF-srDFT models
will depend on how correlation effects can be split into long- and
short-range contributions. This is clear in van der Waals systems in 
which,
for example, MP2-srDFT or CC-srDFT models perform relatively 
well~\cite{srdftnancy,ccsrdft}. In
multi-configurational systems, however, it is in general not possible to
interpret static and dynamical correlations as long- and short-range
ones, respectively. As a result, even though MC-srDFT performs better than regular DFT
in stretched molecules for example~\cite{JCPunivmu}, the approximate
short-range density-functional part of the energy is usually not accurate enough since
it has somehow to describe a part of static
correlation~\cite{JCP10_Andy_RSAC}. In this respect, approximate short-range functionals
that have been developed so far (see Sec.~\ref{sec-compdetails}) need improvements and work is currently
in progress in this direction.
\\
In this paper, the time-dependent
regime will be explored in details only for  
the variational HF-srDFT and MC-srDFT methods. The extension to non-variational
srDFT schemes, which is briefly mentioned in Sec.~\ref{vp_section}, is
left for future work.
\subsection{Variational principle in the time-dependent
regime}\label{vp_section}
The extension of exact srDFT to the time-dependent regime 
could in principle be achieved when considering the
time-dependent Schr\"{o}dinger equation of an auxiliary long-range
interacting system~\cite{JCP12_Pernal_tddmft-srdft}, according to the first Runge and Gross
theorem~\cite{PRL84_RGth}, that is without using a variational
principle. If the long-range interaction is treated approximately, at
the MCSCF level for example, the Ehrenfest
theorem~\cite{jcp85_Aarhus_resp} could then be applied in order to
obtain the time-dependent MC-srDFT
response functions. As an alternative, we explore in this section the
possibility of formulating exact time-dependent srDFT variationally, using the
quasienergy formalism~\cite{ijqc98_Aarhus_resp}. As pointed out in
Ref.~\cite{quasienergy_tddft_Olav}, the latter is arguably more attractive than
the Ehrenfest method in
that it provides a unified framework for applying variational and non-variational
long-range post-HF methods, by analogy with time-independent theory, to
which it naturally reduces in the limit of a static local
potential~\cite{ijqc98_Aarhus_resp}. As
an additional advantage, the permutational
symmetries (with respect to the exchange of perturbation operators) is
manifest in the quasienergy method.\\ 
In order to obtain a time-dependent extension of
Eq.~(\ref{energyminpsi}), we use in the following the recent work of
Vignale~\cite{PRA08_Vignale_TDVP,PRA11_Vignale_TDVP_reply_Schirmer}. 
Let us consider the action integral~\cite{PRL84_RGth} expression
\begin{eqnarray}\label{taquasiener}\begin{array} {l}
{\displaystyle
\mathcal{Q}[{\Psi}]={\displaystyle\int_{t_0}^{t_1} Q[{\Psi}](t)\; dt}
}, 
\\
\\
\displaystyle Q[{\Psi}](t)=\frac{\langle {\Psi}(t)\vert
\hat{T}+\hat{W}_{ee}+\hat{V}(t)-{\rm i}\frac{\partial}{\partial t}\vert
{\Psi}(t)\rangle}{\langle {\Psi}(t)\vert{\Psi}(t)
\rangle},\\
\end{array}
\end{eqnarray}
which is
defined for a given time-dependent wave function
${\Psi}(t)$ 
and a given time-dependent local potential $\hat{V}(t)=\int d\mathbf{r}\;
v(\mathbf{r},t)\,\hat{n}(\mathbf{r})$. The operator $\hat{W}_{ee}$ denotes the regular
two-electron repulsion. 
Note that, in Eq.~(\ref{taquasiener}), ${\Psi}(t)$ is not assumed to be
normalized so that,
when considering infinitesimal variations of the time-dependent quasienergy
$Q[{\Psi}](t)$ in the following, no normalization constrain will be needed. For time-independent local
potentials and wave functions, the time-dependent quasienergy reduces to the usual
energy.
In Floquet theory,
which is considered in Sec.~\ref{subsec_MCtddft}, the action integral
over a period $T$ equals what is usually referred to as a {\it
quasienergy}~\cite{ijqc98_Aarhus_resp,quasienergy_tddft_Olav} 
multiplied by $T$. In the general (non-periodic) case, the quasienergy
would be equal to the action integral divided by $(t_1-t_0)$, which is a
constant. For simplicity, we will refer to
$\mathcal{Q}[{\Psi}]$ as quasienergy even though it is an action and not
an energy,
strictly speaking. 
The time-dependent Schr\"{o}dinger equation 
\begin{eqnarray}\label{TDSeqwithquasiener}\begin{array} {l}
{\displaystyle\left(\hat{T}+\hat{W}_{ee}+\hat{V}(t)-{\rm
i}\frac{\partial}{\partial t}\right)\vert\tilde{\Psi}(t)\rangle={Q[\tilde{\Psi}](t)}\vert\tilde{\Psi}(t)\rangle},\\
\end{array}
\end{eqnarray}
where $\tilde{\Psi}(t_0)$ is assumed to be normalized, is then
equivalent to the variational principle based on the quasienergy 
\begin{eqnarray}\label{TDvarprinc}\begin{array} {l}
{\displaystyle
\delta\mathcal{Q}[\tilde{\Psi}]=-{\rm
i}{\left[\langle\tilde{\Psi}(t)\vert\delta{\Psi}(t)\rangle\right]^{t_1}_{t_0}},
}
\end{array}
\end{eqnarray}
where variations $\tilde{\Psi}(t)\rightarrow
\tilde{\Psi}(t)+\delta\Psi(t)$ around the exact solution
$\tilde{\Psi}(t)$ are considered. 
Using the boundary conditions
$\delta{\Psi}(t_0)=\delta{\Psi}(t_1)=0$, 
the variational principle can be simply written as 
\begin{eqnarray}\label{TDvarprinczero}\begin{array} {l}
{\displaystyle
\delta\mathcal{Q}[\tilde{\Psi}]=0.
}
\end{array}
\end{eqnarray}
This stationary condition is convenient for deriving approximate time-dependent
response properties based on variational methods such as HF and 
MCSCF~\cite{ijqc98_Aarhus_resp}. The time-dependent extension of both
HF-srDFT and MC-srDFT schemes can then be achieved along the same lines,
provided that time-dependent srDFT can be expressed in terms of a variational
principle, which is actually not
trivial~\cite{PRA08_Vignale_TDVP,PRA11_Vignale_TDVP_reply_Schirmer}.
This point is addressed in the rest of this section.\\ 
According to the first Runge and Gross
theorem~\cite{PRL84_RGth}, for a
fixed initial wave function $\tilde{\Psi}(t_0)$, the
following density-functional quasienergy can be defined:
\begin{eqnarray}\label{DFtaquasienerdef}\begin{array} {l}
\mathcal{Q}[n]=\mathcal{B}[n]+
{\displaystyle\int_{t_0}^{t_1} dt\int d{\mathbf r} \;
v(\mathbf{r},t)n(\mathbf{r},t)},
\end{array}
\end{eqnarray}
 where the universal functional    
\begin{eqnarray}\label{DFtaBquasiener}\begin{array} {l}
\mathcal{B}[n]={\displaystyle\int_{t_0}^{t_1} dt\; 
\langle
\tilde{\Psi}[n](t)\vert \hat{T}+\hat{W}_{ee}-{\rm
i}\frac{\partial}{\partial t}\vert
\tilde{\Psi}[n](t)\rangle
}, 
\end{array}
\end{eqnarray}
is calculated with the time-dependent wave function
$\tilde{\Psi}[n](t)$ associated to a fully interacting system
whose time-dependent density equals $n(\mathbf{r},t)$:
\begin{eqnarray}\label{TDSeqwithquasiener-density}\begin{array} {l}
{\displaystyle\left(\hat{T}+\hat{W}_{ee}+\hat{V}[n](t)-{\rm
i}\frac{\partial}{\partial t}\right)\vert\tilde{\Psi}[n](t)\rangle={Q[n](t)}\vert\tilde{\Psi}[n](t)\rangle},\\
\\
\displaystyle
\hat{V}[n](t)=\int
d\mathbf{r}\;v[n](\mathbf{r},t)\;\hat{n}(\mathbf{r}).
\end{array}
\end{eqnarray}
The density $\tilde{n}(\mathbf{r},t)$ associated to the exact solution
$\tilde{\Psi}(t)$ of the time-dependent Schr\"{o}dinger Eq.~(\ref{TDSeqwithquasiener})
can then be obtained when applying the following variational principle 
\begin{eqnarray}\label{TDvarprincdensity}\begin{array} {l}
\displaystyle\delta\mathcal{Q}[\tilde{n}]=-{\rm
i}{\langle\tilde{\Psi}[\tilde{n}](t_1)\vert{\delta\tilde{\Psi}[\tilde{n}](t_1)}\rangle},
\end{array}
\end{eqnarray}
where the boundary condition $\delta\tilde{\Psi}[\tilde{n}](t_0)=0$ was used. As
pointed out by Vignale~\cite{PRA08_Vignale_TDVP}, though the boundary conditions $\delta
n(\mathbf{r},t_0)=\delta n(\mathbf{r},t_1)=0$ are fulfilled, the
condition $\delta\mathcal{Q}[\tilde{n}]=0$ is {\it not}. Indeed, the wave function
$\tilde{\Psi}[\tilde{n}+\delta n](t)=\tilde{\Psi}[\tilde{n}
](t)+\delta\tilde{\Psi}[\tilde{n}](t)$ accumulates the change of the density $\delta
n(\mathbf{r},t)$ over the time interval $[t_0,t_1]$: 
\begin{eqnarray}\label{deltapsint1}
\displaystyle {\delta\tilde{\Psi}[\tilde{n}](t_1)}&=&\int_{t_0}^{t_1} dt\int  d{\mathbf
r} \;\frac{\delta \tilde{\Psi}}{\delta n(\mathbf{r},t)}[\tilde{n}](t_1)\delta
n(\mathbf{r},t)\nonumber\\
&\neq&0,
\end{eqnarray}
so that the variational principle in Eq.~(\ref{TDvarprincdensity}) can
be rewritten as
\begin{eqnarray}\label{TDvarprincdensity2}\begin{array} {l}
\displaystyle\frac{\delta\mathcal{Q}}{\delta n(\mathbf{r},t)}[\tilde{n}]=
-{\rm
i}{\left\langle\tilde{\Psi}[\tilde{n}](t_1)
\middle\vert\frac{\delta\tilde{\Psi}}{\delta
n(\mathbf{r},t)}[\tilde{n}](t_1)\right\rangle}.
\end{array}
\end{eqnarray}
As in the time-independent regime (see Sec.~\ref{sec_srdft}), the universal functional $\mathcal{B}[n]$ defined in
Eq.~(\ref{DFtaBquasiener}) can be split into long-range and short-range
parts:
\begin{eqnarray}\label{RSquasiener}\begin{array} {l}
\mathcal{B}[n]=\mathcal{B}^{\rm lr,\mu}[n]+\mathcal{Q}_{\rm Hxc}^{\rm
sr,\mu}[n],\\
\\
\mathcal{B}^{\rm lr,\mu}[n]= 
\\
\\
 {\displaystyle\int_{t_0}^{t_1} dt}\displaystyle 
 {\langle \tilde{\Psi}^{\mu}[n](t)\vert \hat{T}+\hat{W}^{\rm
lr,\mu}_{ee}-{\rm i}\frac{\partial}{\partial t}\vert
\tilde{\Psi}^{\mu}[n](t)\rangle}
,
\end{array}
\end{eqnarray}
where $\tilde{\Psi}^{\mu}[n](t)$ is the time-dependent wave function associated to a long-range interacting system whose density equals
$n(\mathbf{r},t)$:
\begin{eqnarray}\label{tdSeqlong-rangeint}\begin{array} {l}
{\displaystyle\left(\hat{T}+{\hat{W}^{\rm lr,\mu}_{ee}}+\hat{V}^{\mu}
[n](t)-{\rm i}\frac{\partial}{\partial t}\right)\vert{\tilde{\Psi}^{\mu}[n](t)}\rangle}\\
\\
={Q^{\mu}[n](t)}{\vert\tilde{\Psi}^{\mu}[n](t)\rangle},\\
\\
\displaystyle
\hat{V}^{\mu}[n](t)=\int
d\mathbf{r}\;v^{\mu}[n](\mathbf{r},t)\;\hat{n}(\mathbf{r}).
\end{array}
\end{eqnarray}
Note that the initial state $\tilde{\Psi}^{\mu}[n](t_0)$ is fixed and
equal to the auxiliary long-range interacting wave function
$\tilde{\Psi}^{\mu}(t_0)$ whose density equals the one of the real fully
interacting initial wave function $\tilde{\Psi}(t_0)$.  
The universal long-range functional defined in Eq.~(\ref{RSquasiener})
can be expressed in terms of the auxiliary long-range interacting
time-dependent quasienergy as follows
\begin{eqnarray}\label{lrquasienerfunficti}\begin{array} {l}
\mathcal{B}^{\rm lr,\mu}[n]=\displaystyle\int_{t_0}^{t_1} dt
\;\Big( Q^{\mu}[n](t)\\
\\
\displaystyle
\hspace{2.5cm}- \int d\mathbf{r}\;v^{\mu}[n](\mathbf{r},t)\,
n(\mathbf{r},t)\Big),   
\end{array}
\end{eqnarray}
so that, using the simplified expression
\begin{eqnarray}\label{deltafictiquasi}\begin{array} {l}
\displaystyle\int_{t_0}^{t_1} dt
\;\delta Q^{\mu}[n](t)=-{\rm
i}{\langle\tilde{\Psi}^{\mu}[n](t_1)\vert{\delta\tilde{\Psi}^{\mu}[n](t_1)}\rangle}\\
\\
\displaystyle
\hspace{0.5cm}+ \int_{t_0}^{t_1} dt\int d\mathbf{r}\;\delta v^{\mu}[n](\mathbf{r},t)\,
n(\mathbf{r},t),   
\end{array}
\end{eqnarray}
as well as Eqs.~(\ref{DFtaquasienerdef}) and (\ref{RSquasiener}), the variational principle in Eq.~(\ref{TDvarprincdensity2}) provides the
exact expression for the local potential which reproduces the
time-dependent density $\tilde{n}(\mathbf{r},t)$ of the fully interacting wave function
$\tilde\Psi(t)$ from a long-range interacting one: 
\begin{eqnarray}\label{exacttdvmupot}
\displaystyle
v^{\mu}[\tilde{n}](\mathbf{r},t)&=&v(\mathbf{r},t)+\frac{\delta\mathcal{Q^{\rm sr,\mu}_{\rm
Hxc}}}{\delta n(\mathbf{r},t)}[\tilde{n}]\nonumber\\
\nonumber\\
\displaystyle
&&-{\rm
i}{\left\langle\tilde{\Psi}^{\mu}[\tilde{n}](t_1)
\middle\vert\frac{\delta\tilde{\Psi}^{\mu}}{\delta
n(\mathbf{r},t)}[\tilde{n}](t_1)\right\rangle}\nonumber\\
\nonumber\\
&&+
{\rm
i}{\left\langle\tilde{\Psi}[\tilde{n}](t_1) \middle\vert\frac{\delta\tilde{\Psi}}{\delta
n(\mathbf{r},t)}[\tilde{n}](t_1)\right\rangle}.
\end{eqnarray}
It is important to keep in mind that, as further discussed in Sec.~\ref{sec_interp_poles}, 
the time evolution of the auxiliary
long-range interacting system yields the time-dependent response of the {\it exact} density but {\it
not} the response of the exact
wave function.
In the particular case of a periodic perturbation, which is considered
in the rest of this work, both fully and long-range interacting wave
functions remain unchanged after a period $T$:
\begin{eqnarray}\label{periodicwf}\begin{array} {l}
\tilde{\Psi}[n](t_0+T)=\tilde{\Psi}[n](t_0)=\tilde{\Psi}(t_0),\\
\\
\tilde{\Psi}^{\mu}[n](t_0+T)=\tilde{\Psi}^{\mu}[n](t_0)=\tilde{\Psi}^{\mu}(t_0),
\end{array}
\end{eqnarray}
so that, for $t_1=t_0+T$, the last two terms in the right-hand side of
Eq.~(\ref{exacttdvmupot}) are equal to zero. In addition, if the
adiabatic approximation is used for describing the short-range
interaction, the srHxc density-functional quasienergy expression
is simplified as follows:   
\begin{eqnarray}\label{}\begin{array} {l}
\mathcal{Q}_{\rm Hxc}^{\rm sr,\mu}[n]\;\rightarrow\; \displaystyle\int_{t_0}^{t_1}
dt\; E^{\rm sr,\mu}_{\rm Hxc}[n(t)],  
\end{array}
\end{eqnarray}
where $E^{\rm sr,\mu}_{\rm Hxc}[n]$ is the time-independent srHxc
density-functional introduced in Eq.~(\ref{energyminpsi}), and the time-dependent
potential in Eq.~(\ref{exacttdvmupot}) becomes 
\begin{eqnarray}\label{srpotadiaapprox}\begin{array} {l}
\displaystyle
v^{\mu}[\tilde{n}](\mathbf{r},t)\;\rightarrow\;v(\mathbf{r},t)+\frac{\delta{E^{\rm sr,\mu}_{\rm
Hxc}}}{\delta n(\mathbf{r})}[\tilde{n}(t)].\\
\\
\end{array}
\end{eqnarray}
Combining Eqs.~(\ref{timeindepsrdfteq}), (\ref{tdSeqlong-rangeint}) and (\ref{srpotadiaapprox}),
we conclude that the time-dependent density $\tilde{n}(\mathbf{r},t)$ of the real fully interacting
system can be approximated from an auxiliary long-range interacting one whose wave function
$\tilde{\Psi}^{\mu}(t)$ fulfills, within the short-range adiabatic
approximation,
\begin{eqnarray}\label{Atdsrdfteq}\begin{array} {l}
{\displaystyle\left(\hat{T}+{\hat{W}^{\rm
lr,\mu}_{ee}}+\hat{V}(t)+{\hat{V}^{\rm sr,\mu}_{\rm
Hxc}}[{n_{\tilde{\Psi}^{\mu}(t)}}]-{\rm i}\frac{\partial}{\partial t}\right)\vert{\tilde{\Psi}^{\mu}(t)}\rangle}\\
\\
={Q^{\mu}(t)}{\vert\tilde{\Psi}^{\mu}(t)\rangle},\\
\end{array}
\end{eqnarray}
which is equivalent to the stationary condition
\begin{eqnarray}\label{VPsrdftadia}\begin{array} {l}
\delta\mathcal{Q}^{\mu}[{{\tilde{\Psi}^{\mu}}}]=0,
\end{array}
\end{eqnarray}
where the wave-function-dependent range-separated quasienergy
$\mathcal{Q}^{\mu}[{{{\Psi}}}]$ is defined as
\begin{eqnarray}\label{quasienerAsrdft}\begin{array} {l}
\displaystyle\mathcal{Q}^{\mu}[{{{\Psi}}}]=\int_{t_0}^{t_1}
dt\; \frac{\langle {{\Psi}(t)}\vert \hat{T}+{\hat{W}^{\rm
lr,\mu}_{ee}}+\hat{V}(t)-{\rm i}\frac{\partial}{\partial t}\vert
{{\Psi}(t)}\rangle}{\langle
{{\Psi}(t)}\vert{{\Psi}(t)} \rangle}\\
\\
\displaystyle
\hspace{1.3cm}+\int_{t_0}^{t_1} dt\; {E^{\rm sr,\mu}_{\rm
Hxc}}[{n_{{\Psi}(t)}}]. \\
\end{array}
\end{eqnarray}
Note that
the time-dependent KS equation, as formulated within the adiabatic
approximation, is recovered from Eq.~(\ref{Atdsrdfteq}) when $\mu=0$,
while the $\mu\rightarrow+\infty$
limit corresponds to the time-dependent Schr\"{o}dinger equation. When 
$0<\mu<+\infty$, a rigorous combination of density-functional
and wave function theories is obtained in the time-dependent regime. As shown in the
following, a multi-configuration extension of regular
TD-DFT can then be formulated in this context when describing the long-range
interaction at the MCSCF level. Let us mention that non-variational
methods such as CC could also be merged with TD-DFT, using
Eq.~(\ref{Atdsrdfteq}) in combination with a Lagrangian
formalism~\cite{ijqc98_Aarhus_resp}.


\subsection{Multi-configuration range-separated TD-DFT based on Floquet theory
}\label{subsec_MCtddft}

We work in this section in the framework of Floquet
theory~\cite{ijqc98_Aarhus_resp} where the time-dependent
perturbation is periodic: 
\begin{eqnarray}\label{periodicpert1}\begin{array} {l}
\displaystyle{{\hat{V}}(t)=\hat{V}_{\rm ne}
  +\sum_{x}\sum^N_{k=-N}e^{-{\rm i}\omega_kt}\varepsilon_x(\omega_k)\hat{V}_x},\\
\displaystyle \omega_k=\frac{2\pi k}{T},\\
\\
\displaystyle
\hat{V}_x=\int d\mathbf{r}\; v_x(\mathbf{r})\hat{n}(\mathbf{r}),
\end{array}
\end{eqnarray}
and the quasienergy in Eq.~(\ref{quasienerAsrdft}) is calculated over a period 
$\int_{t_0}^{t_1} dt\longrightarrow \int_0^{T} dt$. 
Since the long-range interaction in 
Eq.~(\ref{Atdsrdfteq}) is treated explicitly, the exact time-dependent
wave function $\tilde{\Psi}^{\mu}(t)$ is a multi-determinant one. As an
approximation, we consider the following MCSCF-type parametrization
consisting of exponential unitary
transformations~\cite{jcp85_Aarhus_resp}: 
\begin{eqnarray}\label{tdmcparametr}\begin{array} {l}
\vert\tilde{\Psi}^{\mu}(t)\rangle=e^{{\rm i}\hat{\kappa}(t)}e^{{\rm
i}\hat{S}(t)}\vert\Psi_0^{\mu}\rangle,
\end{array}
\end{eqnarray}
where $\Psi_0^{\mu}$ denotes the unperturbed time-independent MC-srDFT wave
function and 
\begin{eqnarray}\label{tdrotation}\begin{array} {l}
{\displaystyle
\hat{\kappa}(t)=\sum_{l,i}e^{-{\rm i}\omega_l
t}\kappa_{i}(\omega_l)\hat{q}_i^{\dagger}+e^{-{\rm i}\omega_l
t}\kappa^*_{i}(-\omega_l)\hat{q}_i,
}
\\
\\
{\displaystyle
\hat{S}(t)=\sum_{l,i}e^{-{\rm i}\omega_l
t}S_{i}(\omega_l)\hat{R}_i^{\dagger}+e^{-{\rm i}\omega_l
t}S^*_{i}(-\omega_l)\hat{R}_i.
}
\end{array}
\end{eqnarray}
The singlet excitation and state-transfer operators are defined as follows 
\begin{eqnarray}\label{qidagridag}\begin{array} {l}
{\displaystyle
\hat{q}_i^{\dagger}=\hat{E}_{pq}=\hat{a}^{\dagger}_{p\alpha}\hat{a}_{q\alpha}+\hat{a}^{\dagger}_{p\beta}\hat{a}_{q\beta};
\;\; p>q},\\
\\
\hat{R}_i^{\dagger}=\vert i \rangle\langle \Psi_0^{\mu}\vert. 
\end{array}
\end{eqnarray}
Note that the TD-HF-srDFT scheme is a particular case of
Eq.~(\ref{tdmcparametr}), where the unperturbed MC-srDFT wave function would be
replaced by the HF-srDFT determinant, and orbital rotations only would
be considered. The resulting linear response equations would then be 
formally identical to standard TD-DFT equations based on hybrid
density-functionals~\cite{CP05_Pawel_tddft}. Returning to the
multi-configuration case, the TD-MC-srDFT wave function in
Eq.~(\ref{tdmcparametr}) is fully determined by the
Fourier component vectors
\begin{eqnarray}\label{rspvec}\begin{array} {l}
\Lambda(\omega_l)= \begin{bmatrix}
\kappa_{i}(\omega_l)\\
S_{i}(\omega_l) \\
\kappa^*_{i}(-\omega_l)\\
S^*_{i}(-\omega_l)	  \end{bmatrix},
	  \\
\end{array}
\end{eqnarray}
for which we consider in the following the Taylor expansion through
first order:
\begin{eqnarray}\label{linearrspvec}\begin{array} {l}
$${\displaystyle{\Lambda(\omega_l)}=\sum^N_{k=-N,x}\varepsilon_x(\omega_k){\left
. \frac{\partial {\Lambda(\omega_l)}}{\partial \varepsilon_x(\omega_k)}\right |_{0}}+ \ldots}$$
\end{array}
\end{eqnarray}
Rewriting the variational condition in Eq.~(\ref{VPsrdftadia}) as
follows
\begin{eqnarray}\label{varcondquasisrdft}\begin{array} {l}
{\displaystyle\forall\;\varepsilon_x(\omega_k) \;\;\;\;\;\;\frac{\partial \mathcal{Q}^{\rm \mu}}{\partial
{\Lambda^{\dagger}(-\omega_l)
}}
=0}, 
\end{array}
\end{eqnarray}
the linear response equations are simply obtained by differentiation
with respect to the perturbation strength
$\varepsilon_x(\omega_k)$~\cite{jcp02_Trond_tdhf,CP05_Pawel_tddft}: 
\begin{eqnarray}\label{linearrspsrdft}\begin{array} {l}
\displaystyle \left(\left .
\frac{d}{d\varepsilon_x(\omega_k)}\frac{\partial \mathcal{Q}^{\rm
\mu}}{\partial {\Lambda^{\dagger}(-\omega_l)
}}\right)\right |_{0}=0. 
\end{array}
\end{eqnarray}
According to Eq.~(\ref{quasienerAsrdft}), the quasienergy can be
decomposed 
as follows: 
\begin{eqnarray}\label{decompMCquasiener}\begin{array} {l}
{\mathcal{Q}^{\rm \mu}={\mathcal{Q}^{\rm
lr,\mu}}+{\mathcal{Q}_{\rm Hxc}^{\rm sr,\mu}}},\\
\end{array}
\end{eqnarray}
where the purely long-range MCSCF part equals 
\begin{eqnarray}\label{quasienerlr}\begin{array} {l}
{\mathcal{Q}^{\rm lr, \mu}}={\displaystyle\int_{0}^{T} dt
\;\langle \tilde{\Psi}^{\mu}(t) \,\vert\,
\hat{T}+{\hat{W}^{\rm lr,\mu}_{ee}}+\hat{V}(t)}
\,\vert\, \tilde{\Psi}^{\mu}(t)\rangle\\
{\displaystyle
\hspace{1cm}+\int_{0}^{T} dt\;\langle
\tilde{\Psi}^{\mu}(t)\,\vert\,-{\rm i}\frac{\partial}{\partial t}
\,\vert\,\tilde{\Psi}^{\mu}(t)\rangle},
\end{array}
\end{eqnarray}
and the purely short-range DFT contribution is written as
\begin{eqnarray}\label{quasienersr}\begin{array} {l}
{\mathcal{Q}^{\rm
sr,\mu}}={\displaystyle\int_{0}^{T}{E_{\rm Hxc}^{\rm
sr,\mu}}[{n}_{\tilde{\Psi}^{\mu}(t)}]\; dt},\\
\\
{n}_{\tilde{\Psi}^{\mu}(t)}(\mathbf{r})=\langle\tilde{\Psi}^{\mu}(t)\vert\hat{n}(\mathbf{r})\vert\tilde{\Psi}^{\mu}(t)\rangle.
\end{array}
\end{eqnarray}
The terms arising from the derivatives of the former can be computed with a regular linear response MCSCF
code~\cite{jcp85_Aarhus_resp,jcp88_hjj_lrmcscf_imp}, using long-range two-electron
integrals, and thus do not require additional implementation efforts.
On the other hand, standard TD-DFT codes cannot be used
straightforwardly for computing the srDFT terms since the density is now obtained from a MCSCF-type
wave function instead of a single KS determinant,
terms describing how the density changes when the configuration coefficients change are also needed. 
The srDFT contributions to the linear response equations can be
decomposed as follows   
\begin{eqnarray}\label{derivaQsr}\begin{array} {l}
\displaystyle \left(\left .
\frac{d}{d\varepsilon_x(\omega_k)}\frac{\partial {\mathcal{Q}^{\rm
sr,\mu}}}{\partial {\Lambda^{\dagger}(-\omega_l)}}\right)\right |_{0}=
\frac{d}{d\varepsilon_x(\omega_k)}\frac{\partial }{\partial
{\Lambda^{\dagger}(-\omega_l)}}
\\
\\
\Bigg(\displaystyle
\int_{0}^{T} \langle
\tilde{\Psi}^{\mu}(t) \vert {\hat{V}_{\rm Hxc}^{\rm
sr,\mu}[n_{\Psi^{\mu}_0}]}\vert \tilde{\Psi}^{\mu}(t)\rangle \;dt
\\
\\
\displaystyle
+\frac{T}{2}\sum_{m,n}\delta(\omega_m+\omega_n)
\displaystyle
\int d\mathbf{r}d\mathbf{r'}\;
{{K_{\rm Hxc}^{\rm sr,\mu}}
(\mathbf{r},\mathbf{r'})}\\
\\
\hspace{1.5cm}\times\;\Lambda^{\dagger}(-\omega_m)n^{[1]\mu}(\mathbf{r})n^{[1]\mu\dagger}(\mathbf{r'})\Lambda(\omega_n)
\left.\Bigg)\right |_{0}, 
\end{array}
\end{eqnarray}
where $K_{\rm Hxc}^{\rm sr,\mu}(\mathbf{r},\mathbf{r'})=\delta^2 E_{\rm
Hxc}^{\rm sr,\mu}/\delta n(\mathbf{r})\delta
n(\mathbf{r'})[n_{\Psi^{\mu}_0}]$ denotes the
srHxc kernel calculated for the unperturbed density and the gradient density vector
equals
\begin{eqnarray}\label{gradientdensvector}\begin{array} {l}
n^{[1]\mu}(\mathbf{r})= \begin{bmatrix}
\langle\Psi_0^{\mu}\vert[\hat{q}_i,\hat{n}(\mathbf{r})]\vert\Psi_0^{\mu}\rangle\\
\langle\Psi_0^{\mu}\vert[\hat{R}_i,\hat{n}(\mathbf{r})]\vert\Psi_0^{\mu}\rangle\\
\langle\Psi_0^{\mu}\vert[\hat{q}^{\dagger}_i,\hat{n}(\mathbf{r})]\vert\Psi_0^{\mu}\rangle\\
\langle\Psi_0^{\mu}\vert[\hat{R}^{\dagger}_i,\hat{n}(\mathbf{r})]\vert\Psi_0^{\mu}\rangle\\
\end{bmatrix}
.
\end{array}
\end{eqnarray}
The linear response Eq.~(\ref{linearrspsrdft}) can thus
be rewritten as~\cite{jcp85_Aarhus_resp,jcp88_hjj_lrmcscf_imp} 
\begin{eqnarray}\label{lr_srdft_eq}\begin{array} {l}
\displaystyle
\left . \Bigg(E^{[2]\mu}-\omega_lS^{[2]\mu}\Bigg)\frac{\partial
\Lambda(-\omega_l)}{\partial \varepsilon_x(\omega_k)}\right |_{0} ={\rm i}
V_x^{[1]\mu}\delta(\omega_k+\omega_l),
\end{array}
\end{eqnarray}
where the Hessian is split as follows:
\begin{eqnarray}\label{srdfthessian}\begin{array} {l}
E^{[2]\mu}=E_0^{[2]\mu}+K_{\rm Hxc}^{\rm sr,\mu}.
\end{array}
\end{eqnarray}
The MCSCF-type Hessian $E_0^{[2]\mu}$ is based on the auxiliary long-range interacting
Hamiltonian $\hat{T}+{\hat{W}^{\rm lr,\mu}_{ee}}+\hat{V}_{\rm
ne}+{\hat{V}_{\rm Hxc}^{\rm sr,\mu}[n_{\Psi^{\mu}_0}]}$, that is used as $H_0$ in Eqs.~(9) and (10) of
Ref.~\cite{jcp88_hjj_lrmcscf_imp}, and the srHxc kernel
contribution is defined as
\begin{eqnarray}\label{srkernelhessian}\begin{array} {l}
\displaystyle
K_{\rm Hxc}^{\rm sr,\mu}=\int d\mathbf{r}d\mathbf{r'}\; K_{\rm Hxc}^{\rm
sr,\mu}(\mathbf{r},\mathbf{r'})n^{[1]\mu}(\mathbf{r})n^{[1]\mu\dagger}(\mathbf{r'}).
\end{array}
\end{eqnarray}
Both $E_0^{[2]\mu}$ and $S^{[2]\mu}$ matrices (see Eqs.~(7) and (8) in
Ref.~\cite{jcp88_hjj_lrmcscf_imp}) are built from the unperturbed
MC-srDFT wave function $\Psi^{\mu}_0$ and the gradient property vector
$V_x^{[1]\mu}$ equals $\int d\mathbf{r}\;
v_x(\mathbf{r})n^{[1]\mu}(\mathbf{r})$. 
Excitation energies can thus be calculated at the MC-srDFT level when 
solving iteratively~\cite{jcp88_hjj_lrmcscf_imp}
\begin{eqnarray}\label{MCCasidaeq}\begin{array} {l}
\displaystyle
\Big(E^{[2]\mu}-\omega S^{[2]\mu}\Big)X(\omega)=0,
\end{array}
\end{eqnarray}
which can be considered a multi-configuration extension of the Casida
equations~\cite{Casida_eqs}. 

\subsection{Interpretation of the TD-MC-srDFT
poles}\label{sec_interp_poles}
The
linear response Eq.~(\ref{MCCasidaeq}) computes the poles of the TD-MC-srDFT wave
function. As already mentioned in Sec.~\ref{vp_section}, a solution can
only be
interpreted as an excitation energy if it is a pole of the
TD-MC-srDFT density. In this respect, the double excitation recovered in
the $\mu=0$ limit of a simple two-state TD-MC-srDFT model (see Appendix)
is not physical. It is in principle less problematic for non-zero $\mu$
values, even small ones, since the ground-state MC-srDFT
wave function becomes multi-determinantal. This is analyzed 
further in Secs.~\ref{subsec_SigmaG} and \ref{subsec_Be}. In order to obtain a smoother connection
to regular TD-DFT when $\mu\rightarrow 0$,  
an effective orbital rotation vector 
\begin{eqnarray}\label{effrspvec}\begin{array} {l}
\tilde{\mathcal{K}}(\omega_l)= \begin{bmatrix}
\tilde{\kappa}_{i}(\omega_l)\\
\tilde{\kappa}^{*}_{i}(-\omega_l)\\
\end{bmatrix},
	  \\
\end{array}
\end{eqnarray}
such that
\begin{eqnarray}\label{effrotdef}\begin{array} {l}
\tilde{n}^{[1]\mu\dagger}(\mathbf{r})\tilde{\mathcal{K}}(\omega_l)= 
n^{[1]\mu\dagger}(\mathbf{r})\Lambda(\omega_l), \\
\\
\tilde{n}^{[1]\mu\dagger}(\mathbf{r})=
 \begin{bmatrix}
\langle\Psi_0^{\mu}\vert[\hat{q}_i,\hat{n}(\mathbf{r})]\vert\Psi_0^{\mu}\rangle\\
\langle\Psi_0^{\mu}\vert[\hat{q}^{\dagger}_i,\hat{n}(\mathbf{r})]\vert\Psi_0^{\mu}\rangle\\
\end{bmatrix},
\end{array}
\end{eqnarray}
could be introduced and the TD-MC-srDFT linear response
Eq.~(\ref{lr_srdft_eq}) reformulated
in terms of $\left .\partial\tilde{\mathcal{K}}(-\omega_l)/\partial
\varepsilon_x(\omega_k)\right |_{0}$. Work is currently in progress in
this direction.

\section{Computational details}\label{sec-compdetails}

The TD-MC-srDFT linear response Eq.~(\ref{lr_srdft_eq}) has been
implemented in a development version of the {\tt DALTON2011} program
package~\cite{daltonpack}. Calculations have been performed with spin-independent short-range
functionals, considering both local density (srLDA) and
generalized gradient (srGGA) approximations. In the former case, we used the
srLDA functional of Toulouse, Savin and Flad~\cite{savinbook,toulda}.
We used, as srGGA functional, the Perdew-Burke-Ernzerhof-type (PBE) functional of
Goll, Werner and Stoll~\cite{ccsrdft}, which is denoted srPBEgws in the
following. Basis sets are
aug-cc-pVQZ~\cite{augQZbe} for both H$_2$ and Be systems. 
We furthermore performed all-electron TD-MC-srDFT linear response calculations 
of ferrocene Fe(C$_5$H$_5$)$_2$\ (iron \emph{bis}-cyclopentadienyl,
FeCp$_2$) examining excitation energies and oscillator strengths of the lowest singlet excited \emph{d-d}\ and
ligand-to-metal charge transfer states.
The geometrical parameters for the eclipsed FeCp$_2$\ conformer
(D$_{5h}$\ symmetry), compiled in Table S1 in the Supplementary
Material~\cite{supp_mat_td-mc-srdft},  
were taken from the recent work by Coriani \emph{et al.} \cite{fecp2_geo_cc2} who carried out highly accurate geometry optimizations at the CCSD(T) level yielding 
close agreement with experiment and other available \emph{ab initio}\ data \cite{fecp2_geo_cc1, fecp2_geo_caspt2, fecp2_exc_sacci, fecp2_exc_tddft3}.  
We employed triple-$\zeta$\ cc-pVTZ-DK basis sets for all elements \cite{dunning89, cc-pVTZ-DK-3d}\ where scalar-relativistic effects were taken into account by means 
of the Douglas-Kroll-Hess second-order (DKH2) Hamiltonian. 
We did not account for spin-orbit effects 
which were found by Scuppa and co-workers \cite{fecp2_exc_tddft3} to be
of only minor importance for the singlet excitation spectrum of ferrocene.
The excited state manifold of FeCp$_2$\ was computed with the regular TD-HF and TD-MCSCF approaches as well as 
the combined TD-HF-srLDA, TD-HF-srPBEgws and TD-MC-srPBEgws models. 
The initial CASSCF optimization step was performed for both MC-srDFT and regular MCSCF using the well-established CAS(10,10) 
active space with \emph{10} electrons in \emph{10} orbitals \cite{fecp2_geo_caspt2},
comprising the Cp-ligand $\pi$\ orbitals in addition to the Fe $3d4s$\ shells.
For analysis purposes, we carried out standard 
scalar-relativistic DKH2 TD-DFT calculations using pure 
LDA~\cite{dft-Vosko-CJP1980a} and PBE~\cite{dft-Perdew-PRL1996a}
functionals, as well as the hybrid Becke three-parameter Lee-Yang-Parr
B3LYP~\cite{dft-Becke-JCP1993a} and the Coulomb attenuated method CAM-B3LYP~\cite{dft-Yanai-CPL2004a}\ 
functionals, within the adiabatic approximation. 
The $\mu$ parameter was set in all TD-HF-srDFT and TD-MC-srDFT calculations to $0.4$\ a.u.~unless otherwise specified. 
This value relates to a prescription given in
Refs.~\cite{JCPunivmu,JCPunivmu2} for an optimal treatment of
short-range electron correlation in ground-state MC-srDFT calculations and should, in principle, be re-considered in
the time-dependent regime. As illustrated in Sec.~\ref{subsec_Be}, the choice of $\mu$
in TD-MC-srDFT calculations is important since it affects excitation
energies significantly. This will be further analyzed in future work.

\section{Results and discussion}

\subsection{$^1\Sigma^+_u$ excited states of H$_2$}\label{subsec_SigmaU} 

Our TD-MC-srDFT calculations have been performed for H$_2$ with the minimal
$1\sigma_g1\sigma_u$ active space which however can recover
$^1\Sigma^+_u$ excitation energies very close
to the FCI ones already at the TD-MCSCF level (see Fig.~\ref{h2_sigmaU}~(a)).
Potential curves obtained with the
srLDA and srPBEgws functionals were found to be on top of each other.
Therefore only the formers are discussed in the following. As shown in
Fig.~\ref{h2_sigmaU}~(b), the TD-HF-srLDA and TD-MC-srLDA
methods give, near the equilibrium H--H distance (about 1.4 a.u.), the same excitation energies for the first four
$^1\Sigma^+_u$ states when $\mu$ is set to 0.4 a.u. This is due to the
fact that (i) the ground-state HF-srLDA and MC-srLDA wave functions are,
in this case,
almost identical~\cite{JCPunivmu} (ii) these excitations all correspond, predominantly, to single
excitations from the $1\sigma_g$ to $\sigma_u$
orbitals~\cite{JCP12_Baerends_tddmft}. The latters are indeed well
described by orbital rotations, like in standard TD-DFT, with no need
for a long-range multi-configuration treatment. Note that the excitation energies are, in this case, more
accurate than the pure TD-LDA ones: using the FCI curves as reference, the error is approximately divided
by two. Upon bond stretching, the difference between TD-HF-srLDA and
TD-MC-srLDA excitation energies becomes 
significant for the $4^1\Sigma^+_u$ state already around 2.5 a.u.
Indeed, while the former is associated to the singly excited
configuration $(1\sigma_g)^1(4\sigma_u)^1$, the latter corresponds to the
doubly excited configuration $(1\sigma_u)^1(2\sigma_g)^1$. This double
excitation is obtained as a single excitation applied to the doubly excited
configuration $(1\sigma_u)^2$ whose weight 
in the ground-state MC-srLDA wave function is less than 1\% but not
strictly equal to zero. For the same bond distance, the $3^1\Sigma^+_u$ state, which is dominated by
the singly excited configuration $(1\sigma_g)^1(3\sigma_u)^1$, acquires
also, but to a less pronounced extent than for $4^1\Sigma^+_u$, a doubly excited
character $[(1\sigma_g)^2\rightarrow(1\sigma_u)^1(2\sigma_g)^1]$. The $1^1\Sigma^+_u$ and
$2^1\Sigma^+_u$ states are mainly combinations of the singly excited
configurations $(1\sigma_g)^1(1\sigma_u)^1$ and
$(1\sigma_g)^1(2\sigma_u)^1$, which explains the smaller difference,
compared to the $4^1\Sigma^+_u$ state, between the TD-HF-srLDA and
TD-MC-srLDA excitation energies. Around 5 a.u., the $4^1\Sigma^+_u$
state obtained at the TD-MC-srLDA level is dominated by the
doubly excited configuration $(1\sigma_u)^1(3\sigma_g)^1$ while, at the
regular TD-MCSCF level, both $(1\sigma_u)^1(3\sigma_g)^1$ and
$(1\sigma_u)^1(2\sigma_g)^1$ configurations are important. This
difference could be justified by the fact that long-range interactions
only are described within the MCSCF model while short-range interactions
are assigned to DFT. In other words, the ground-state wave function and its linear
response are not expected to be the same at the MC-srDFT and regular MCSCF
levels. As
shown in Sec.~\ref{vp_section}, this should be expected only for the
densities. The large underestimation of the $4^1\Sigma^+_u$ excitation energy, by
about 0.1 a.u., is mainly due to the approximate (adiabatic srLDA) potential and
kernel used. 
It is important to notice that the explicit treatment, at the MCSCF
level, of the long-range interaction enables a multi-configuration
description, within a TD-DFT framework, of the excited states upon bond
stretching. This is illustrated by the increasing doubly excited
character $[(1\sigma_g)^2\rightarrow(1\sigma_u)^1(2\sigma_g)^1]$ of the
$3^1\Sigma^+_u$ state when enlarging the bond distance from 2.5 to 5
a.u. This explains the increasing difference between the TD-HF-srLDA and
TD-MC-srLDA excitation energies. For the latter, the doubly excited
character is induced by the single excitation
$[(1\sigma_u)^2\rightarrow(1\sigma_u)^1(2\sigma_g)^1]$ applied, in the
long-range MCSCF linear response calculation,
to the configuration $(1\sigma_u)^2$ whose weight increases in the
ground-state MC-srLDA wave function (0.6\% at 2.5 a.u. and 23\% at 5 a.u.). 
Let us mention that, even though TD-HF-srLDA and TD-LDA methods cannot
describe double excitations, they provide at 5 a.u. a more accurate 
$3^1\Sigma^+_u$ excitation energy than the TD-MC-srLDA
approach, an example of fortuitous error cancellation.
Upon further stretching, both TD-LDA and TD-HF-srLDA
potential curves indeed deviate increasingly from the FCI one, while the 
TD-MC-srLDA curve remains parallel to the latter. 
Such a fortuitous error cancellation does not happen for the 
$2^1\Sigma^+_u$ and $1^1\Sigma^+_u$ states, and the TD-MC-srLDA
performs better than TD-LDA and TD-HF-srLDA methods
upon bond stretching. At 5 a.u., the $2^1\Sigma^+_u$ state is dominated
by both
singly excited $(1\sigma_g)^1(2\sigma_u)^1$ and doubly excited
$(1\sigma_u)^1(2\sigma_g)^1$ configurations for which the MC-srLDA
linear response coefficients are equal to 0.33 and 0.13
in absolute value, respectively. On the other hand, the $1^1\Sigma^+_u$ state is dominated by the
singly excited $(1\sigma_g)^1(1\sigma_u)^1$ ionic configuration which is
included in the active space and thus described, not by orbital
rotations, but by configuration rotations instead. As
expected~\cite{JCP12_Baerends_tddmft},
the TD-HF-srLDA $1^1\Sigma^+_u$ potential curve does not exhibit a
minimum, exactly as with TD-LDA, simply because the $1\sigma_g$ and
$1\sigma_u$ orbitals become degenerate in the dissociation limit. On the
contrary, the TD-MC-srLDA curve has a minimum. 
This clearly shows that describing the long-range part of the
electron-electron interaction at the MCSCF level in a range-separated TD-DFT framework
can improve remarkably the
exchange-correlation kernel used in standard TD-DFT calculations.
Note also that the TD-DMFT-srLDA ($\mu=0.7$) $1^1\Sigma^+_u$ excitation
energy of Pernal~\cite{JCP12_Pernal_tddmft-srdft} (3.02 eV), which was
computed for
a bond distance of $10.0~a.u.$, is largely 
improved at the TD-MC-srLDA ($\mu=0.7$) level
(8.45 eV) when compared to FCI (9.90 eV). All these results were
obtained with the aug-cc-pVTZ~\cite{augQZbe} basis set. Even though we
use a different srLDA functional, we can reasonably conclude that
the large errors obtained at the TD-DMFT-srLDA level
are essentially due to the long-range adiabatic BB approximation.
Returning to the TD-MC-srLDA ($\mu=0.4$)
results, the 
$1^1\Sigma^+_u$ excitation energy is still significantly underestimated for
large bond distances. 
The srLDA potential and kernel should then be improved in order to obtain
more accurate results. 
One can also
notice that the FCI $1^1\Sigma^+_u$ potential curve minimum (around 4
a.u.) is shifted to about 5.5 a.u. at the TD-MC-srLDA level. Indeed, up to bond
distances of 4 a.u., there are no significant differences between
TD-HF-srLDA and TD-MC-srLDA excitation energies. This is due to the
relatively small $\mu=0.4$ value, which ensures that, at the equilibrium
distance, most of the
electron correlation is assigned to the short-range interaction and thus
treated in DFT. The assignment of static
correlation to the long-range interaction appears clearly in the dissociation
limit, that is when both $1\sigma_g$ and $1\sigma_u$ natural
orbitals are singly occupied in the ground-state MC-srLDA wave function~\cite{JCPunivmu}. For intermediate bond distances, the
range-separation of static and dynamic correlations is not clear
anymore~\cite{JCP10_Andy_RSAC}. It
means that the short-range potential and kernel should be accurate
enough to obtain reliable excitation energies for distances where a part
of static
correlation is inevitably assigned to the short-range interaction. The
situation is of course less critical for larger $\mu$ values but, in
this case, more electron correlation must be described by the MCSCF which is
not appealing, in terms of computational cost, for larger scale calculations. Using a
multi-configuration short-range exact exchange energy
expression~\cite{TousrXmd,PaolasrXmd,PaolasrXmd_prb}
while keeping $\mu=0.4$~a.u. is a possible alternative currently under
investigation. Such a scheme would largely reduce self-interaction
errors~\cite{JCPunivmu} in the ground-state MC-srLDA energy and is expected to affect 
excitation energies through the improved short-range potential and kernel.

\subsection{$^1\Sigma^+_g$ excited states of H$_2$}\label{subsec_SigmaG} 

The TD-MC-srLDA method was also applied to the calculation of the first four
$^1\Sigma^+_g$ states of H$_2$, setting $\mu$ to 0.4 a.u. and using the
minimal $1\sigma_g1\sigma_u$ active space. The latter enables to
recover, at the TD-MCSCF level, excitation energies which are rather close to the FCI ones as shown in
Fig.~\ref{h2_sigmaG} (a). Near the equilibrium distance the
$n^1\Sigma^+_g$ excited states ($n=2,\ldots,5$), as described by the
TD-MCSCF model, are dominated by the singly excited configuration
$(1\sigma_g)^1(n\sigma_g)^1$. This statement holds at the
TD-MC-srLDA level for the first three lowest $^1\Sigma^+_g$ excited
states, which explains why, in the light of Sec.~\ref{subsec_SigmaU}, the corresponding excitation energies are very
close to the TD-HF-srLDA ones, as shown in Fig.~\ref{h2_sigmaG} (b). On the other hand, the TD-MC-srLDA
$5^1\Sigma^+_g$ state is dominated by the doubly excited configuration
$(1\sigma_u)^2$ already at the equilibrium distance. This is illustrated
by the large difference between the TD-HF-srLDA and TD-MC-srLDA
excitation energies at 1.4 a.u.
Note that, at the TD-MCSCF level, the doubly excited character of the
$5^1\Sigma^+_g$ state appears only at 2 a.u., when the slope of the
potential curve suddenly changes.
Upon
bond stretching, the avoided crossing of the TD-MC-srLDA
$5^1\Sigma^+_g$ and $4^1\Sigma^+_g$ states occurs around 2 a.u. The
latter gains a doubly excited character which explains the deviation of
its excitation energy from the TD-HF-srLDA one.
Note that regular TD-LDA as
well as TD-HF-srLDA methods completely miss the $5^1\Sigma^+_g$ state
for stretched geometries.
Returning to the $4^1\Sigma^+_g$ state, the TD-MC-srLDA exhibits an
avoided crossing with the $3^1\Sigma^+_g$ state around 2.5 a.u. The
latter is then dominated by the doubly excited configuration
$(1\sigma_u)^2$. As a consequence the TD-HF-srLDA and TD-MC-srLDA
$3^1\Sigma^+_g$ excitation energies differ significantly when stretching
the bond beyond 2.5 a.u. Another avoided crossing occurs, at 3 a.u. in
the TD-MC-srLDA model,
between the $3^1\Sigma^+_g$ and $2^1\Sigma^+_g$ states. While regular
TD-LDA as well as TD-HF-srLDA keep on describing the lowest
$^1\Sigma^+_g$ excited state as singly excited for large bond distances,
the TD-MC-srLDA method is able to catch the double excitation
$(1\sigma_g)^2\rightarrow(1\sigma_u)^2$, which is included in the
active space. For analysis purposes, TD-MC-srLDA excitation energies
have been computed when $\mu=0$ (see Fig.~\ref{h2_sigmaG_mu0.0}).
As shown in the
Appendix, even though the
ground-state MC-srLDA wave function reduces to the regular
KS-LDA determinant, the TD-MC-srLDA $(\mu=0)$ method recovers not only
the TD-LDA spectrum but also the double excitation
$(1\sigma_g)^2\rightarrow(1\sigma_u)^2$
whose excitation energy is found to be twice the KS orbital energy difference
$2(\varepsilon_{\sigma_u}-\varepsilon_{\sigma_g})$. This explains
the crossings in Fig.~\ref{h2_sigmaG_mu0.0}. As pointed out in
Sec.~\ref{sec_interp_poles}, the double excitation should in principle be
disregarded since it does not correspond to a pole of the density. Still, it is
interesting to note that the TD-MC-srLDA ($\mu=0.4$) avoided crossings shown in Fig.~\ref{h2_sigmaG}
(b) originate from this unphysical solution. Let us stress that, for
$\mu=0.4$, the double excitation described within TD-MC-srLDA is, on the
other hand, physical. This is ensured by non-zero orbital rotation coefficients (not shown)
in the corresponding linear response vector as well as the multi-configurational
character of the ground-state MC-srLDA wave function in stretched
geometries. 
Finally, as in Sec.~\ref{subsec_SigmaU}, we again conclude that
the inaccuracy of the TD-MC-srLDA ($\mu$=0.4)
method is essentially due to the approximate short-range potential and 
kernel used.

\subsection{Singlet excited states of the beryllium
atom}\label{subsec_Be}

Singlet excitation
energies of the beryllium atom have been computed at the
TD-MC-srDFT level, using the srPBEgws functional and the $2s2p$ active
space, when varying the $\mu$ parameter. Results are shown in
Figs.~\ref{XEmu_be} (a) and \ref{XEmu_be} (b). Comparison is made with the linear response
TD-CCSD results which are used as reference. Let us first mention that,
in the $\mu=0$ limit of the TD-MC-srPBEgws model, standard TD-PBE excitation energies are recovered.
Interestingly the $1^1D$ state, which corresponds to the double
excitation $(2s)^2\rightarrow(2p)^2$ and which is absent in the standard
TD-PBE spectrum, is also described at the TD-MC-srPBEgws level even when
$\mu=0$. In this particular case, the excitation energy actually equals
twice the KS-PBE orbital energy difference
$2(\varepsilon_{2p}-\varepsilon_{2s})$. As shown in the Appendix, this
is due to the fact that the double excitation is included in the
active space and can, therefore, be treated explicitly. In this case,
the excitation energy turns out to be equal the TD-CCSD one.
Nevertheless, when $\mu$ is strictly equal to zero, this double
excitation is unphysical as mentioned in Sec.~\ref{sec_interp_poles}.
For non-zero $\mu$ values, even small ones, this is not the case
anymore since the ground-state MC-srPBEgws wave function is
multi-configurational and orbital rotations appear in the linear response vector. Let us
mention that, at
the TD-MC-srLDA level, we obtained a $1^1D$ excitation energy of 7.17 eV
when $\mu=0.1$ which corresponds to an absolute error of 0.02 eV. In the
TD-DMFT-srLDA ($\mu=0.1$) method of Pernal~\cite{JCP12_Pernal_tddmft-srdft},
the absolute error is much larger (2 eV). 
(We note that Pernal used a different srLDA functional than we used, 
namely the one by Paziani {\it et al.}~\cite{PaolasrXmd_prb},
but we do not expect this to be important for the absolute error.) 
As pointed out by the author this error
might be due to the approximate density-matrix functional used for
describing the long-range interaction while, in the TD-MC-srDFT model,
the latter is explicitly treated at the MCSCF level. When increasing the $\mu$
parameter, the TD-MC-srPBEgws $1^1D$ excitation energy deviates slightly
from the TD-CCSD one, by less than 0.15 eV. Note that the
TD-MC-srLDA $1^1D$ excitation energy was found to be equal
to 6.96 eV for $\mu=0.4$~a.u., which corresponds to an 
absolute deviation of 0.23 eV. 
The absolute error obtained at the TD-DMFT-srLDA ($\mu=0.4$) level by
Pernal~\cite{JCP12_Pernal_tddmft-srdft} is much larger (3.38 eV),
which we again attribute
to the approximate density-matrix functional used for modeling the
long-range interaction. The other excitation
energies are all underestimated at the regular TD-PBE level. Those can
be improved significantly when $\mu$ becomes larger than 0.2 a.u. The
difference between the TD-HF-srPBEgws and TD-MC-srPBEgws excitation
energies increases then with $\mu$ since electron correlation is
transferred from the short-range DFT part to the long-range MCSCF part
of the linear response equations. We note
that, for most of the states, TD-MC-srPBEgws can perform better than
TD-HF-srPBEgws and the regular TD-MCSCF model, which is recovered in the $\mu\rightarrow+\infty$
limit, but different $\mu$ values
should then be used. As extensively discussed in Ref.~\cite{JCPunivmu} it is
more appealing, in order to have a general method, to set the $\mu$ parameter to a fixed value, using
prescriptions for an optimal treatment of correlation for example, and
improve the accuracy of the short-range functionals. Work is in progress
in this direction. 

\subsection{ferrocene}\label{subsec_ferrocene}

A major objective stimulating the development of the TD-MC-srDFT model is to pave the way for an efficient 
computational tool of predictive power that allows us to study excitation spectra of large, complex molecular systems where for example local and charge-transfer (CT) excitations often co-occur. 
As an illustrative example we computed the low-lying singlet excitation spectrum of ferrocene for which  
experimental solution data \cite{fecp2_exp_1967,fecp2_exp_1971}\ as well as ample theoretical \emph{ab initio}\ 
gas-phase reference values  \cite{fecp2_exc_sacci, fecp2_exc_tddft1, fecp2_exc_tddft2, fecp2_exc_tddft3, fecp2_exc_tddft4} are available. 

A ground-state HF calculation for ferrocene with the Cp-rings in D$_{5h}$ symmetric configuration around the $d^6$\ Fe(II) center yields the following energetic order of iron-centered high-lying occupied (non-bonding) and anti-bonding LUMO $d$-orbitals:   
$$ a_{1}^{'} (d_{z^2}; d_{\sigma}) \approx e_{2}^{'} (d_{x^{2}-y^{2}}, d_{xy}; d_{\delta}) \ll e_1^{''} (d_{xz}, d_{yz}; d_{\pi}).$$ 
In addition, we find as highest occupied MOs (doubly degenerate HOMO orbitals) predominantly Cp-centered, bonding $\pi$-orbitals. In summary, the valence configuration reads as:
$ (a_{1}^{'})^2\ (e_{2}^{'})^4\ (e_1^{''})^4\ (e_1^{''})^0 $.

Based on this ground state occupation we can expect 
a manifold of three doubly-degenerate (E$^{''}_1$, E$^{''}_2$, E$^{''}_1$) singlet metal-centered valence transitions which will be examined in Section \ref{d-d-trans}\ as well as ligand-to-metal CT excitations in the ultra-violet visible spectrum 
to be discussed in Section \ref{l-m-trans}. In particular in the latter case it will be interesting to examine whether our new TD-MC-srDFT scheme can improve TD-DFT/CAM-B3LYP. 
In TD-DFT/CAM-B3LYP both long- and short-range correlations are treated in DFT 
whereas in TD-MC-srDFT the long-range correlation is handled with MCSCF. 
 
All experimental data available to us is recorded in solution, 
and a fair comparison of the computed data 
would thus require to account for solvent effects which was beyond the scope of the present investigation.  
Work in that direction is currently in progress \cite{hedegaard_2012}. 
Hence, we \emph{assume} that the metal-centered transitions are rather insensitive to solvent effects 
due to their local character whereas the CT transitions are more likely to be sensitive 
to solvation effects because of potentially strong ligand-solvent interactions. 

Before embarking on details of the excited state spectrum compiled in Table \ref{fecp2-comparison-d-d}\ 
we commence with a brief examination of the geometry dependence of our data.  
For this purpose, we performed TD-DFT/CAM-B3LYP calculations with
ferrocene (\emph{i}) in a staggered conformation with D$_{5d}$ symmetry
(for geometrical data see Table S2 in the Supplementary
Material~\cite{supp_mat_td-mc-srdft}) 
and (\emph{ii}) in a second eclipsed conformation with geometrical parameters taken from experiment \cite{fecp2_exp_struct1,fecp2_exp_struct2,fecp2_exp_struct3} which was used 
by Ishimura \emph{et al.}~\cite{fecp2_exc_sacci} in their symmetry
adapted cluster-configuration interaction (SACCI) excited state
calculations (see Table S3 in the Supplementary
Material~\cite{supp_mat_td-mc-srdft}). 
We find that the lowest-lying singlet excited states are hardly affected by adapting to a staggered conformation. 
The $d$-$d$\ excitation energies are lowered to a minor degree  
to 2.29 eV, 2.62 eV and 3.43 eV with a maximum deviation of $\Delta_{max}$ = -0.04 eV from the 
reference TD-DFT/CAM-B3LYP values (D$_{5h}$-symmetric configuration) listed in Table \ref{fecp2-comparison-d-d}. 
Larger deviations with $\Delta_{max}$ = -0.10 eV are observed for the second 
eclipsed configuration of ferrocene where the $d$-$d$\ excitation energies read as 2.23 eV, 2.56 eV and 3.37 eV. 
The latter should be taken into account when comparing our final results with the SACCI data given in Table \ref{fecp2-comparison-d-d}. 

\subsubsection{Metal-centered $d$-$d$\ transitions}\label{d-d-trans}

Table \ref{fecp2-comparison-d-d} comprises in the upper part the
spin-allowed singlet $d$-$d$\ transition energies. 
For further analysis we summarize in Table \ref{fecp2-comparison-d-d-delta}\ the relative energy 
gaps $\Delta_1$\ and $\Delta_2$, respectively which are determined from the excitation energy 
differences between E$_2^{''}$\ and the lower E$_1^{''}$\ state and the higher-lying singlet $d$-$d$\ transition 
E$_1^{''}$\ and E$_2^{''}$. 

Turning first to the transition energies computed at the TD-HF level of theory reveals strikingly
the importance of accounting for electron correlation. We find deviations from experiment 
by more than 1.8 eV for the lower E$^{''}_1$\ and E$^{''}_2$ states. The comparatively close match of the 
second E$^{''}_1$\ transition of 3.39 eV with the experimental reference
of 3.82 eV is fortuitous.
The TD-MCSCF approach in contrast allows to recover a significant part of the dynamic electron correlation in the excited states 
since all $d$-orbitals involved in the low-lying $d$-$d$\ transitions are included in the active CAS(10,10) space. 
This indeed leads to overall, greatly improved absolute excitation energies. The third excited state E$^{''}_1$, however, 
is located at only 3.15 eV giving rise to a too close gap $\Delta_2 = 0.19$\ eV and in addition, $\Delta_1$\ hardly changes compared to TD-HF. 

The excited state data derived from regular TD-DFT calculations listed in Tables \ref{fecp2-comparison-d-d}\ and \ref{fecp2-comparison-d-d-delta}\ follow general tendencies 
that can be summarized as follows. 
Introducing a fraction of full-range exact HF exchange like in B3LYP (20 \%) and CAM-B3LYP (19 \%) 
compared to the LDA and PBE functionals correlates with decreasing excitation energies in contrast to 
increasing energetic gaps $\Delta_{1,2}$. In general, this results in a closer agreement with experiment 
using hybrid functionals. Hence, we may order the accuracy of the functionals studied here as 
CAM-B3LYP $\approx$\ B3LYP $>$\ PBE $>$\ LDA. As one could expect for the rather spatially localized $d$-$d$\ transitions\ the hybrid B3LYP functional performs similarly well compared to the more sophisticated 
 CAM-B3LYP functional which includes an additional 46 \% of long-range HF exchange. 
Let us now consider the excitation energies and their energetic
separations $\Delta_1$\ and $\Delta_2$\ computed by means of TD-HF-srDFT.
Using 100\% of long-range HF exchange, irrespective of the choice of
srLDA or srPBE functionals, 
reduces the transition energies and enlarges their relative gaps  $\Delta_1$\ and $\Delta_2$\ in comparison to the regular TD-DFT/LDA and TD-DFT/PBE data, respectively. This is not only in line with 
the findings for (CAM-)B3LYP relative to LDA and PBE but results also in
a good agreement of TD-HF-srDFT with the hybrid functionals. Clearly, $\mu = 0.4$\ a.u.~ is large enough to introduce a 
non-negligible part of HF exchange, since long-range effects do not appear to be significant in the $d$-$d$\ excitations. 
The fact that TD-MC-srPBEgws and TD-HF-srPBEgws results are not equal
elucidates that a non-negligible part of the two-electron interaction is
indeed assigned to the long-range interaction and thus treated by the MCSCF approach within the CAS(10,10) active space. 
In accordance with Fromager \emph{et al.}\ \cite{JCPunivmu},
total energy differences on the order of $10^{-2}$ a.u.~between the
ground-state 
HF-srPBEgws and MC-srPBEgws wave functions, respectively, corroborate our conclusion. 
In passing, we note that our TD-MC-srPBEgws data are in very good agreement with experiment 
with respect to both excitation energies and the relative energetic 
spacings $\Delta_{1,2}$\ of the three lowest electronic singlet $d$-$d$\
transitions. The largest deviation to experiment is for the third
excited state (located at 3.82 eV) 
where all approaches studied here yield too low transition energies except for SACCI which, not taking into account further geometry effects (\emph{vide supra}), slightly overshoots by +0.21 eV. 
 In summary, based on the experimental references, 
 our new TD-MC-srPBEgws method shows significantly enhanced performance compared 
 to the multi-reference wave function SACCI approach which yields only 
 a moderate agreement with experiment both with respect to transition energies and their energetic separation $\Delta_{1,2}$. 
 Moreover, TD-MC-srPBEgws provides a valuable alternative to the otherwise 
 well-performing hybrid (CAM-)B3LYP functionals as well as the model potential LB94 which 
 was explored in the recent work by Scuppa and co-workers \cite{fecp2_exc_tddft3}.

\subsubsection{Ligand $\rightarrow$\ Metal charge-transfer excitations}\label{l-m-trans}

We report in the lower part of Table \ref{fecp2-comparison-d-d}\ CT excitation energies 
computed at the TD-HF, TD-DFT level employing LDA, GGA, B3LYP and CAM-B3LYP functionals along
with our TD-HF/MC-srDFT results. The experimental data refers to the
maximum peak positions in the measured absorption spectrum of ferrocene.
It is readily seen that the TD-HF CT excitation energies are unacceptably high whereas 
the TD-DFT/LDA and TD-DFT/PBE transitions are too close to the lower end of the visible spectrum. 
With the hybrid B3LYP functional, on the other hand, deviations from experiment are reduced to -0.41 and -0.48 eV 
for the first and second CT transitions, respectively, yielding excitation energies of 5.41 eV and 5.72 eV. 
This clearly indicates the importance of using an exact exchange energy, or at least a fraction of it, with regard to 
a more general description of CT states where the extent of exact exchange determines the \emph{partial} correction of 
the ``long-range" self-interaction error intrinsically present for pure LDA and GGA-type functionals. 
The appropriate combination with a suitable correlation functional then leads to satisfying results as seen here for B3LYP. 
Bearing this in mind, the superior performance of CAM-B3LYP compared to B3LYP with respect to the ligand-to-metal CT excitations might not be unexpected 
as CAM-B3LYP comprises in addition to the 19\% of full-range HF exchange energy (B3LYP: 20\%), 46\% of the long-range exact exchange energy. Although only 
a fraction of long-range HF exchange is used, fortuitous error cancellations in combination with an approximate correlation density-functional 
may lead to excellent results, where CAM-B3LYP excitation energies at the triple-$\zeta$\ basis set level nearly 
coincide with the experimental solution data. 
Turning to the oscillator strengths, they are
too weak in all our calculations, which is probably a combination of basis set effects and the solvent
effects on the experiments.\\
A closer inspection of the CT transitions in Table \ref{fecp2-comparison-d-d}\ further reveals 
that both TD-DFT/LDA and TD-HF predict the CT transitions in reversed order.
This feature remains also in the picture predicted by our TD-MCSCF response calculations where 
the description of long-range and short-range dynamic correlation within the CAS(10,10)\ space vastly improves the excitation energies, 
lowering them to 5.50 eV for the A$_2^{''}$\ and 6.24 eV for the E$_1^{'}$\ state while, at large, their relative gap (in reversed order) enlarges to 0.74 eV compared to 0.38 eV in experiment. 
The extended active orbital space treated within the SACCI approach, on the other hand, seems to capture most of the important contributions to the CT transitions, 
thus illustrating the importance of short-range dynamic correlation. Peaks are predicted closely spaced at 6.34 eV and 6.43 eV, respectively, though. 
Similar conclusions can be drawn for the TD-HF-srDFT model with either
the srLDA or srPBEgws functional. The straightforward inclusion of short-range dynamic correlation by means of the srDFT 
functional, using $\mu = 0.4$\ a.u., suffices in this context to recover
the correct CT state order from the $\mu \rightarrow +\infty$\ pure
wave-function-based TD-HF limit. 
In contrast to the local $d$-$d$\ transitions (see
Sec.~\ref{d-d-trans})\ we observe for the first CT state a large difference between the TD-HF-srDFT and 
TD-DFT/CAM-B3LYP approaches which we ascribe mainly to different fractions of long-range HF exchange, 
which amounts to only 65\% in CAM-B3LYP, and more importantly, to missing long-range correlation in TD-HF-srDFT.    
The impact of describing the latter correlation type explicitly, within the MCSCF model, 
is highlighted by our TD-MC-srPBEgws results. It lowers the first
E$_1^{'}$\ transition by 0.35 eV compared to TD-HF-srPBEgws 
while the second main CT peak is pushed upwards to 7.04 eV, thus overshooting the experimental reference by $\approx 0.8$ eV. 
A detailed analysis shows that the second CT transition is, within both
the TD-MCSCF and TD-MC-srPBEgws approaches, fully described 
by orbital rotations whereas the optically allowed E$_1^{'}$\ transition
is largely obtained by configuration rotations. 
However, one cannot make any final conclusions about the quality of the
TD-MC-srPBEgws results compared to experiments
before the solvent effects have been included in the calculations.
 Work on this is currently in progress.
Computing high-level gas phase CC excitation energies would also be
interesting for benchmarking.\\
To sum up, our new proposed TD-MC-srDFT model has promising potential to accurately describe not only local excitations 
as shown for example for the $d$-$d$\ transitions in
Sec.~\ref{d-d-trans}\ but also to simultaneously model charge transfer
excitations and their corresponding relative oscillator strengths. 
It exhibits in particular improved performance compared to both TD-DFT
based on pure density-functionals and TD-MCSCF. In contrast to the
latter approach, the srPBEgws functional combined with a long-range
TD-MCSCF treatment is able to 
describe short-range dynamic correlations and enables to get the two CT states in predicted order. 
The relative energy difference between the two CT states seems too large, though, when compared to experimental data.  
This might be due to self-interactions errors induced by the employed short-range functionals where the formers could be reduced 
choosing a different separation of short-range exchange and correlation energies. Work is currently in progress in this direction. Moreover, 
the inclusion of solvation effects by means of a polarizable embedded framework into the TD-MC-srDFT and TD-DFT response approaches \cite{hedegaard_2012} 
of the {\tt DALTON2011} program framework will facilitate future comparison to experimental works recorded in solution. 
Finally, we note that the TD-MC-srDFT scheme seems to be a viable
alternative to TD-DFT/CAM-B3LYP where, in the former case, long-range correlation effects are 
described at the MCSCF level. Further calibration studies focusing
particularly on such a direct comparison are being prepared. 

\section{Conclusions}\label{conclu}

The extension of multi-determinant range-separated DFT to the
time-dependent regime
has been investigated. Following
Vignale~\cite{PRA08_Vignale_TDVP,PRA11_Vignale_TDVP_reply_Schirmer}, an exact variational
formulation was obtained. Various approximate schemes can then be
formulated, depending on the choice of the post-HF method that is used for describing the long-range
interaction, and the choice of the exchange-correlation functional used
for treating the short-range
interaction. In this work, the combination of long-range MCSCF with
short-range adiabatic LDA and GGA was considered. The corresponding
linear response scheme, referred to as TD-MC-srDFT,  
was then derived within Floquet theory. 
Numerical results obtained for the singlet excited
states of the 
stretched H$_2$ molecule and Be show that TD-MC-srDFT can, in 
contrast to regular TD-DFT, describe
double excitations even though the adiabatic approximation is used for 
the short-range exchange-correlation density-functional contributions. This is
made possible by the MCSCF-type parametrization of the wave function
where, in addition to orbital rotations, configuration rotations
inside the active space can be treated explicitly. For the commonly used
$\mu=0.4$ range-separation parameter value, excitation energies are, in
most cases, much better described with TD-MC-srDFT than  
regular TD-DFT. Huge improvements are also observed, when comparing with
the TD-DMFT-srLDA results of Pernal~\cite{JCP12_Pernal_tddmft-srdft}, in the description
of the $1^1D$
doubly-excited state of Be and the $1^1\Sigma^+_u$ state of the stretched
H$_2$ molecule. Still, in the latter case, 
excitation energies are significantly underestimated. The error is due to
the approximate short-range LDA and GGA potential and kernel that are used.
Increasing $\mu$ improves on the accuracy but then some part of the
short-range electron correlation is transferred from DFT to the MCSCF.
This is of course not appealing, in terms of computational cost, for larger scale calculations.
As an alternative, a different decomposition of the short-range exchange
and correlation energies~\cite{TousrXmd,PaolasrXmd,PaolasrXmd_prb} could be used while keeping the  
$\mu$ parameter set to $0.4$~a.u. Work is currently in progress in this
direction. The accuracy of the TD-MC-srDFT approach using a short-range GGA functional 
was further examined for the low-lying singlet excited states of the 
$d^6$\ metallocene ferrocene. Excitation energies as well as energetic state separations are 
in overall good agreement with experiment for both valence and charge-transfer excitations apart from a single outlier. 
TD-MC-srDFT generally outperforms traditional wave-function and TD-DFT approaches with exception of TD-DFT/CAM-B3LYP that yields 
comparable results with a better performance for the charge-transfer states.   
The latter functional was specifically designed to remedy known shortcomings of TD-DFT \cite{dft-Yanai-CPL2004a} for 
this excitation class, though. Extensive works with the overall aim to shed further light on weaknesses and strengths of our proposed TD-MC-srDFT approach 
with respect to TD-DFT/CAM-B3LYP are therefore in progress. Particular focus is in this context laid 
on model excited-state charge-transfer compounds such as peptides, carotenoides \cite{hedegaard_peptide_2012}\ 
and transition metal complexes \cite{knecht_tmo4x_2012}.

\begin{acknowledgments}
S.K.~acknowledges the Danish Natural Science Research Council for an individual postdoctoral grant 
(10-082944). Computing resources were provided by the Danish Center for Scientific
Computing (DCSC) at the University of Southern Denmark
in Odense. E.F. thanks ANR (contract DYQUMA) as well as Kenneth Ruud,
Radovan Bast, Andreas Thorvaldsen and Julien Toulouse for fruitful discussions on
response theory. 
\end{acknowledgments}

\renewcommand{\theequation}{A\arabic{equation}}
    \setcounter{equation}{0}  
      \section*{APPENDIX: TWO-STATE TD-MC-srDFT MODEL IN THE
	$\mu=0$ LIMIT}  



Let us consider a minimal active space consisting of two Slater
determinants $\vert a^2\rangle$ and $\vert b^2\rangle$ representing
doubly-occupied $\phi_a$ and $\phi_b$ orbitals, respectively. The former determinant corresponds to the ground-state
KS determinant which is recovered at the MC-srDFT level in the $\mu=0$
limit (see Sec.~\ref{sec_srdft}). For simplicity, orbital rotations are not considered
in the following. According to Eqs.~(7) and (8) in
Ref.~\cite{jcp88_hjj_lrmcscf_imp}, the long-range Hessian $E_0^{[2]\mu}$
and the metric $S^{[2]\mu}$ introduced in
Eqs.~(\ref{lr_srdft_eq}) and (\ref{srdfthessian}) become, when $\mu=0$,  
\begin{eqnarray}\label{lrhessianmuzero}\begin{array} {l}
E_0^{[2]0}= \begin{bmatrix}
A^0&B^0\\
B^{0*}&A^{0*} \\
\end{bmatrix}
,\hspace{0.4cm}
S^{[2]0}= \begin{bmatrix}
\Sigma^0&\Delta^0\\
-\Delta^{0*}&-\Sigma^{0*}\\
\end{bmatrix},

\end{array}
\end{eqnarray}
where
\begin{eqnarray}\label{Amuzero}\begin{array} {l}
A^{0}= \begin{bmatrix}
\langle a^2 \vert [\hat{R}_{a^2},[\hat{H}^{\rm
KS},\hat{R}^{\dagger}_{a^2}]]\vert a^2\rangle 
& 
\langle a^2 \vert [\hat{R}_{a^2},[\hat{H}^{\rm
KS},\hat{R}^{\dagger}_{b^2}]]\vert a^2\rangle 
\\
\langle a^2 \vert [\hat{R}_{b^2},[\hat{H}^{\rm
KS},\hat{R}^{\dagger}_{a^2}]]\vert a^2\rangle 
&
\langle a^2 \vert [\hat{R}_{b^2},[\hat{H}^{\rm
KS},\hat{R}^{\dagger}_{b^2}]]\vert a^2\rangle 
\\
\end{bmatrix}
,\\
\\
B^{0}= \begin{bmatrix}
\langle a^2 \vert [\hat{R}_{a^2},[\hat{H}^{\rm
KS},\hat{R}_{a^2}]]\vert a^2\rangle 
& 
\langle a^2 \vert [\hat{R}_{a^2},[\hat{H}^{\rm
KS},\hat{R}_{b^2}]]\vert a^2\rangle 
\\
\langle a^2 \vert [\hat{R}_{b^2},[\hat{H}^{\rm
KS},\hat{R}_{a^2}]]\vert a^2\rangle 
&
\langle a^2 \vert [\hat{R}_{b^2},[\hat{H}^{\rm
KS},\hat{R}_{b^2}]]\vert a^2\rangle 
\\
\end{bmatrix}
,
\\
\\
\Sigma^{0}= \begin{bmatrix}
\langle a^2 \vert [\hat{R}_{a^2}
,\hat{R}^{\dagger}_{a^2}]\vert a^2\rangle 
& 
\langle a^2 \vert [\hat{R}_{a^2}
,\hat{R}^{\dagger}_{b^2}]\vert a^2\rangle 
\\
\langle a^2 \vert [\hat{R}_{b^2}
,\hat{R}^{\dagger}_{a^2}]\vert a^2\rangle 
&
\langle a^2 \vert [\hat{R}_{b^2}
,\hat{R}^{\dagger}_{b^2}]\vert a^2\rangle 
\\
\end{bmatrix},
\\
\\
\Delta^{0}= \begin{bmatrix}
\langle a^2 \vert [\hat{R}_{a^2}
,\hat{R}_{a^2}]\vert a^2\rangle 
& 
\langle a^2 \vert [\hat{R}_{a^2}
,\hat{R}_{b^2}]\vert a^2\rangle 
\\
\langle a^2 \vert [\hat{R}_{b^2}
,\hat{R}_{a^2}]\vert a^2\rangle 
&
\langle a^2 \vert [\hat{R}_{b^2}
,\hat{R}_{b^2}]\vert a^2\rangle 
\\
\end{bmatrix}
,
\\
\\
\hat{R}^{\dagger}_{a^2}=\vert a^2\rangle\langle a^2\vert,\hspace{0.2cm}
\hat{R}^{\dagger}_{b^2}=\vert b^2\rangle\langle a^2\vert,
\end{array}
\end{eqnarray}
and $\hat{H}^{\rm KS}=\hat{T}+\hat{V}_{\rm ne}+{\hat{V}_{\rm
Hxc}}[n_{a^2}]$ denotes the non-interacting KS Hamiltonian. Since
$\hat{H}^{\rm KS}\vert a^2\rangle=2\varepsilon_a\vert a^2\rangle$ and
$\hat{H}^{\rm KS}\vert b^2\rangle=2\varepsilon_b\vert b^2\rangle$, the
matrices in Eq.~(\ref{Amuzero}) can be simplified as follows
\begin{eqnarray}\label{Amuzerosimplified}\begin{array} {l}
A^{0}= \begin{bmatrix}
0 & 0\\
0 & 2(\varepsilon_b-\varepsilon_a)
\end{bmatrix}
,\\
\\
B^{0}=\Delta^{0}=0,\\
\\
\Sigma^{0}=\begin{bmatrix}
0 & 0\\
0 & 1
\end{bmatrix}.
\end{array}
\end{eqnarray}
In addition, the gradient density vector defined in
Eq.~(\ref{gradientdensvector}) becomes in the $\mu=0$ limit:
\begin{eqnarray}\label{gradientdensvectormuzero}\begin{array} {l}
n^{[1]0}(\mathbf{r})= 
\begin{bmatrix}
0\\ 
\langle b^2\vert\hat{n}(\mathbf{r})\vert a^2\rangle\\
0\\
-\langle a^2\vert\hat{n}(\mathbf{r})\vert b^2\rangle\\

\end{bmatrix}
=0,
\end{array}
\end{eqnarray}
since $\vert a^2\rangle$ and $\vert b^2\rangle$ differ by a
double excitation. As a result the kernel contribution to the Hessian defined in
Eq.~(\ref{srkernelhessian}) equals zero when $\mu=0$ so that the total
Hessian is simply $E_0^{[2]0}$. We thus conclude from
Eqs.~(\ref{lrhessianmuzero}) and (\ref{Amuzerosimplified}) that the double
excitation $a^2\rightarrow b^2$ is described within the TD-MC-srDFT model
in the $\mu=0$ limit and the corresponding excitation energy is 
twice the KS orbital energy difference $2(\varepsilon_b-\varepsilon_a)$.
In this respect, the TD-MC-srDFT model does not reduce to standard
TD-DFT, as the TD-HF-srDFT model does, in the $\mu=0$ limit. Still, in
this example, the double excitation is a pole of the TD-MC-srDFT wave
function but not of the density since the linear response contribution to
the density $n^{[1]0\dagger}(\mathbf{r})X(\omega)$ always equals zero, according
to Eq.~(\ref{gradientdensvectormuzero}). From this point of view, the
double excitation obtained at the TD-MC-srDFT ($\mu=0$) level is unphysical.

\clearpage



\newcommand{\THone}[0]{TH} \newcommand{\Aa}[0]{Aa}

\clearpage

\textbf{FIGURE CAPTIONS}

\begin{figure}
\caption{\label{h2_sigmaU} Fromager et al, Journal of Chemical Physics}
\vspace{-1.8cm}
\begin{center}
\begin{tabular}{c}
\hspace{3cm}\resizebox{15cm}{!}{\includegraphics{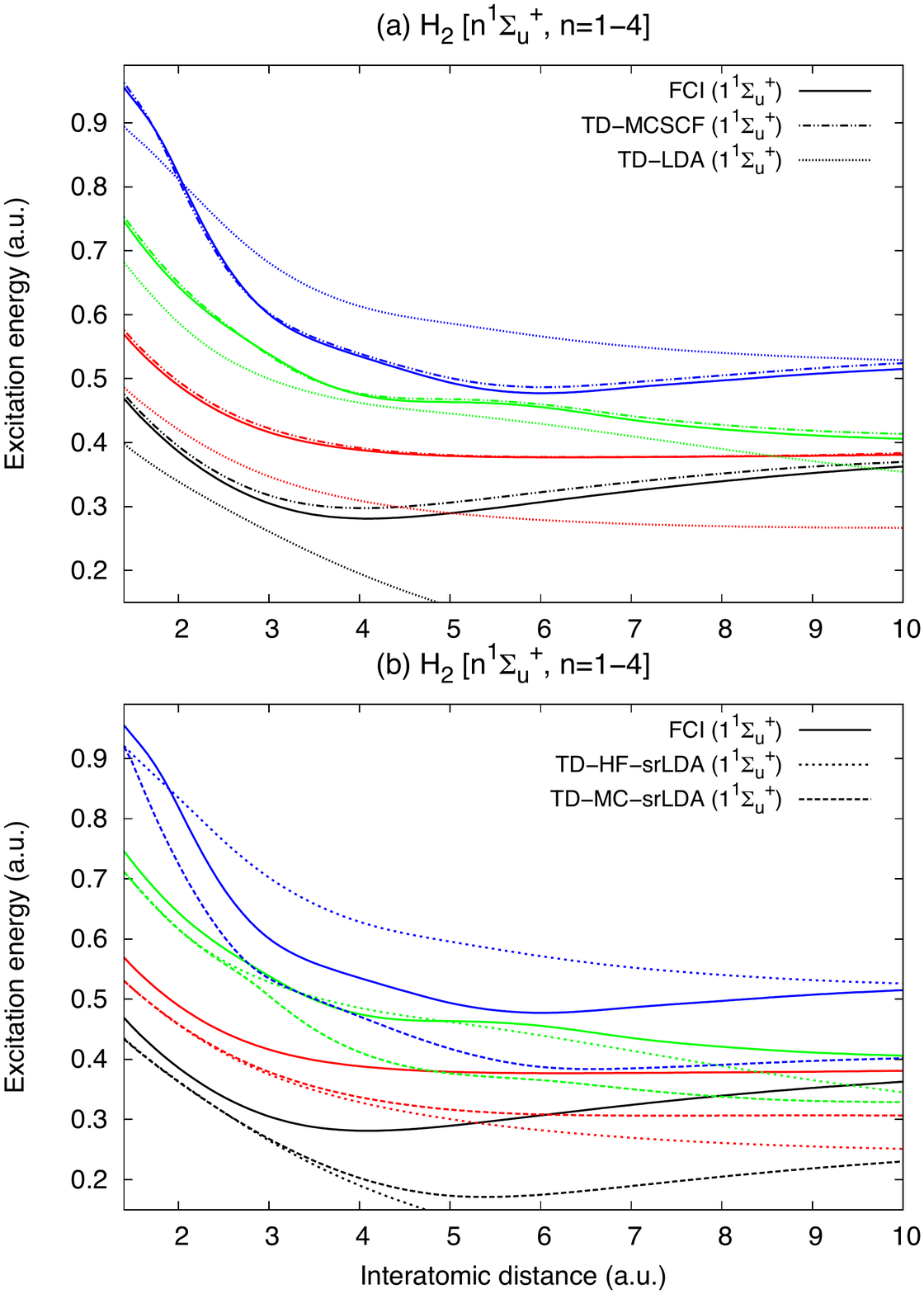}}
\end{tabular}
\end{center}
\end{figure}

\begin{figure}
\caption{\label{h2_sigmaG} Fromager et al, Journal of Chemical Physics}
\begin{center}
\vspace{-0.8cm}
\begin{tabular}{c}
\hspace{3cm}\resizebox{15cm}{!}{\includegraphics{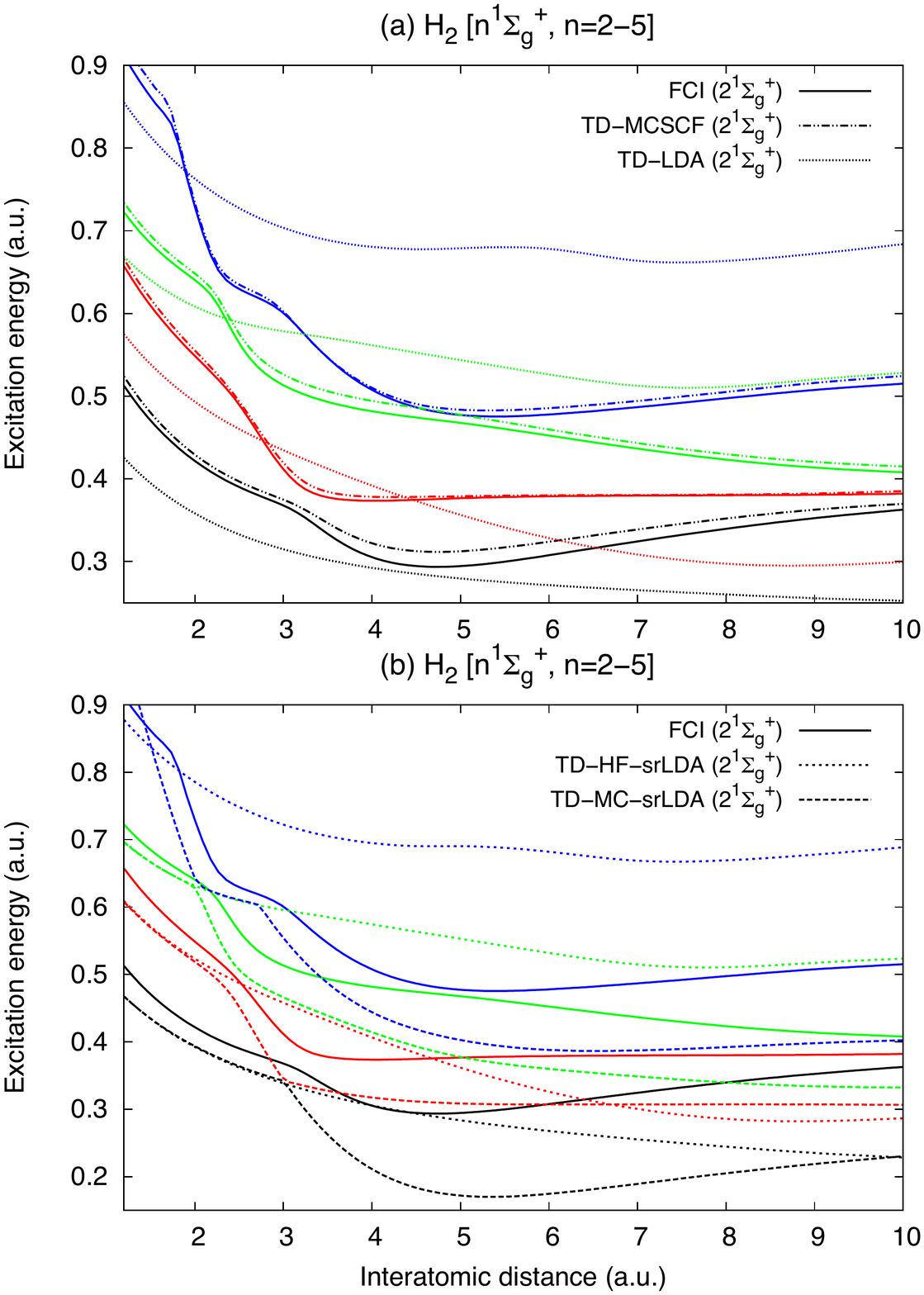}}
\end{tabular}
\end{center}
\end{figure}

\begin{figure}
\caption{\label{h2_sigmaG_mu0.0} Fromager et al, Journal of Chemical Physics}
\begin{center}
\begin{tabular}{c}
\hspace{-1.5cm}\resizebox{16cm}{!}{\includegraphics{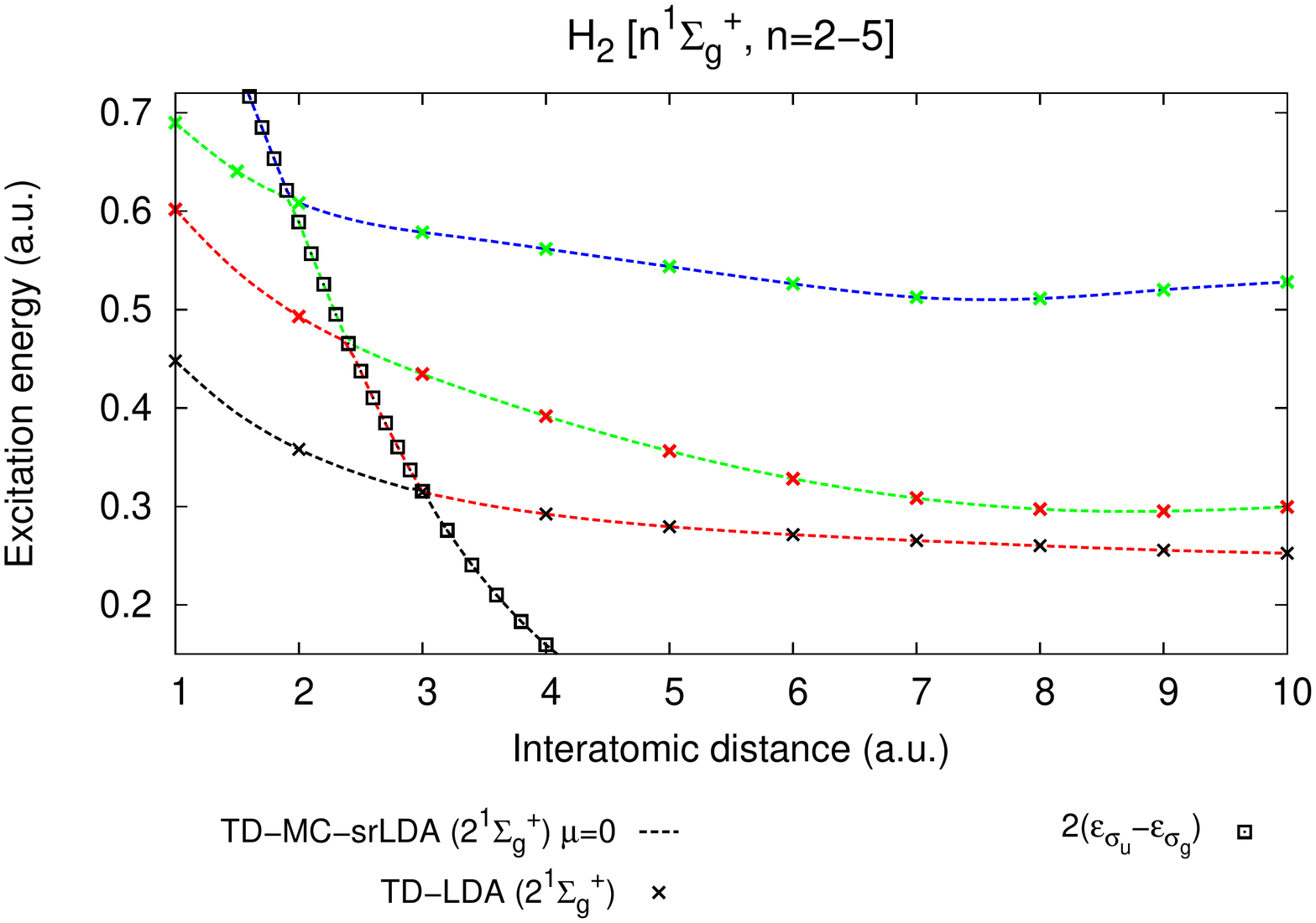}}
\end{tabular}
\end{center}
\end{figure}

\begin{figure}
\caption{\label{XEmu_be} Fromager et al, Journal of Chemical Physics}
\begin{center}
\vspace{-0.5cm}
\begin{tabular}{c}
\resizebox{16cm}{!}{\includegraphics{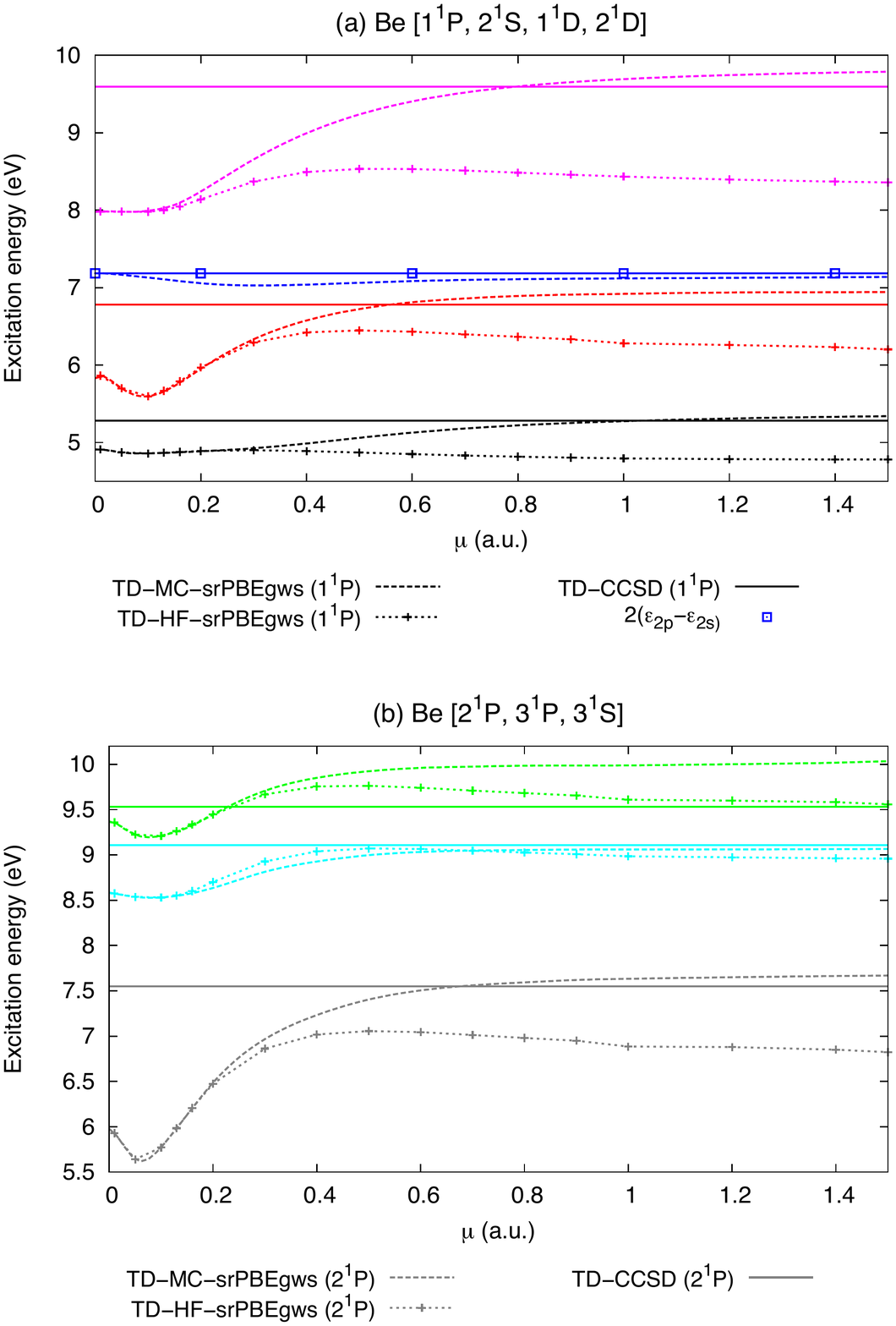}}
\end{tabular}
\end{center}
\end{figure}

\begin{description}

\item[Figure \ref{h2_sigmaU}:] First (black), second (red), third
(green) and fourth (blue) $^1\Sigma^+_u$ excitation
energies in H$_2$ along the bond breaking coordinate calculated with
(a) standard
TD-MCSCF and TD-LDA methods (b) TD-HF-srLDA and TD-MC-srLDA
schemes. Comparison is made with FCI results. Each type of line corresponds to a given method. The $\mu$ parameter
was set to 0.4 a.u. The minimal active space $1\sigma_g1\sigma_u$ was
used. Basis set is aug-cc-pVQZ. 
 

\item[Figure \ref{h2_sigmaG}:] 
First (black), second (red), third
(green) and fourth (blue) $^1\Sigma^+_g$ excitation
energies in H$_2$ along the bond breaking coordinate calculated with
(a) standard
TD-MCSCF and TD-LDA methods (b) TD-HF-srLDA and TD-MC-srLDA
schemes. Comparison is made with FCI results. Each type of line corresponds
to a given method. The $\mu$ parameter
was set to 0.4 a.u. The minimal active space $1\sigma_g1\sigma_u$ was
used. Basis set is aug-cc-pVQZ. 

\item[Figure \ref{h2_sigmaG_mu0.0}:] 

First (black), second (red), third
(green) and fourth (blue) $^1\Sigma^+_g$ excitation
energies in H$_2$ along the bond breaking coordinate obtained with the
TD-MC-srLDA method, setting $\mu=0$, and compared to standard TD-LDA
results. For analysis purposes, twice the $1\sigma_u$ and $1\sigma_g$
KS-LDA orbital energy difference is also plotted. Each type of line or
point corresponds to a given method. 
The minimal active space $1\sigma_g1\sigma_u$ was used. Basis set is aug-cc-pVQZ.

\item[Figure \ref{XEmu_be}:] Excitation energies obtained in Be when varying $\mu$ at 
the TD-HF-srPBEgws and TD-MC-srPBEgws levels for (a) the $1^1P$ (black),
$2^1S$ (red), $1^1D$ (blue) and $2^1D$ (purple) singlet states (b)
the $2^1P$ (grey), $3^1P$ (cyan) and $3^1S$ (green) singlet
states. Comparison
is made with TD-CCSD results. Each type of line corresponds to a given
method. Twice the KS-PBE orbital energy
difference $2(\varepsilon_{2p}-\varepsilon_{2s})$ is also plotted for analysis purposes. The minimal active
space $2s2p$ was used. Basis set is aug-cc-pVQZ.    
\end{description}

\clearpage

\clearpage

\clearpage

\clearpage

\clearpage

\textbf{TABLE CAPTIONS}

\begin{description}


\item[Table \ref{fecp2-comparison-d-d}:] Key spin-allowed singlet Fe $d$-$d$ and lowest optically active singlet transition energies (eV) in ferrocene.
Oscillator strength \emph{f}\ are reported in the row below the charge-transfer state excitation energies for selected, exemplary cases. 
The abbreviation "TD-" has been omitted for compactness in the table header. 

\item[Table \ref{fecp2-comparison-d-d-delta}:] Energy gaps (in eV)
between the lowest three spin-allowed singlet Fe $d$-$d$ transition
energies in ferrocene obtained from TD-MC-srDFT calculations, 
and compared with other \emph{ab initio}\ and experimental data. $\Delta_1$\ refers to the energetic difference between the lower  E$_1^{''}$\ and E$_2^{''}$\ states while 
$\Delta_2$\ quantifies the difference in transition energies between the
upper E$_1^{''}$ and E$_2^{''}$\ states, respectively. The abbreviation
"TD-" has been omitted for compactness in the table header. 
\end{description}

\clearpage

\clearpage

\newpage

\begin{sidewaystable}[h]
\caption{\label{fecp2-comparison-d-d} Fromager et al, Journal of Chemical Physics}
\begin{center}
\begin{tabular}{l@{\hspace{12pt}}c@{\hspace{2pt}}c@{\hspace{2pt}}c@{\hspace{12pt}}c@{\hspace{2pt}}c@{\hspace{2pt}}c@{\hspace{2pt}}c@{\hspace{2pt}}c@{\hspace{12pt}}c@{\hspace{2pt}}c@{\hspace{2pt}}c@{\hspace{12pt}}c}\hline \hline
              &      &             &            &      &             &           &      & CAM-     & HF-    & HF-       & MC-      &  \\
State 
              & HF   & MCSCF       & SACCI$^a$  & LDA  & PBE         & LB94$^b$  & B3LYP&  B3LYP   &  srLDA & srPBE$^c$ &  srPBE$^c$  & Experiment$^{d}$ \\ \hline
 \multicolumn{13}{c}{\emph{$d$-$d$ excitations}}\\
 E$_1^{''}$   & 0.92 & 2.82        & 2.11       & 2.94 & 2.90        &  2.81     & 2.43 & 2.33        &     2.45 &         2.42 &         2.59  & 2.70\\ 
 E$_2^{''}$   & 1.09 & 2.96        & 2.27       & 2.99 & 3.03        &  2.91     & 2.70 & 2.65        &     2.74 &         2.79 &         2.88  & 2.98 \\
 E$_1^{''}$   & 3.39 & 3.15        & 4.03       & 3.52 & 3.60        &  3.44     & 3.48 & 3.45        &     3.29 &         3.41 &         3.50  & 3.82 \\ 
 \multicolumn{13}{c}{\emph{charge-transfer excitations}}\\
  E$_1^{'}$   & 8.49 & 6.24        & 6.34       & 5.13 & 5.09        & n/a       & 5.41 & 5.85        &     6.28 &         6.24 &         5.89  & 5.82$^e$ \\ 
\quad\emph{f} &    - & \emph{0.040}& -          & -    & \emph{0.013}& -         &    - & \emph{0.018}&       -  & \emph{0.024} & \emph{0.024}  & \emph{0.105} \\
  A$_2^{''}$  & 8.12 &        5.50 & 6.43       & 4.89 & 5.46        & n/a       & 5.72 & 6.25        &     6.34 &         6.39 &        7.04   & 6.20$^f$ \\
\quad\emph{f} &   -  & \emph{0.188}& -          & -    & \emph{0.002}& -         &    - & \emph{0.107}& -        & \emph{0.149} & \emph{0.171}  & \emph{0.730} \\ \hline \hline
\multicolumn{13}{l}{$^a$\ Ref. \citenum{fecp2_exc_sacci}, $^b$\ Ref. \citenum{fecp2_exc_tddft3},$^c$\ srPBEgws, 
$^{d}$\ Ref. \citenum{fecp2_exp_1971}, $^e$ value from Ref. \citenum{fecp2_exp_1967}, $^f$\ Ref. \citenum{fecp2_exp_1967}: 6.31 eV (\emph{f}: \emph{0.691}).}
\end{tabular}
\end{center}
\end{sidewaystable}

\clearpage

\begin{sidewaystable}[h]
\caption{\label{fecp2-comparison-d-d-delta} Fromager et al, Journal of Chemical Physics}
\begin{center}
\begin{tabular}{c@{\hspace{12pt}}c@{\hspace{2pt}}c@{\hspace{2pt}}c@{\hspace{12pt}}c@{\hspace{2pt}}c@{\hspace{2pt}}c@{\hspace{2pt}}c@{\hspace{2pt}}c@{\hspace{12pt}}c@{\hspace{2pt}}c@{\hspace{2pt}}c@{\hspace{12pt}}c}\hline \hline
  &    &   &    &    &    &    &   & CAM-  & HF-  & HF-  & MC-  &   \\
  Energy gap & HF   & MCSCF & SACCI$^a$ & LDA  & PBE  & LB94$^b$  & B3LYP & B3LYP & srLDA & srPBE$^c$ & srPBE$^c$ & Experiment$^{d}$ \\ \hline
 $\Delta_1$       & 0.17 & 0.14  & 0.16      & 0.05 & 0.13 &  0.10     & 0.27  & 0.32      & 0.29     & 0.37         & 0.29         & 0.28 \\ 
 $\Delta_2$       & 2.30 & 0.19  & 1.76      & 0.53 & 0.57 &  0.53     & 0.78  & 0.80      & 0.55     & 0.62         & 0.62         & 0.84 \\ \hline \hline
\multicolumn{13}{l}{$^a$\ Ref. \citenum{fecp2_exc_sacci}, $^b$\ Ref. \citenum{fecp2_exc_tddft3}, $^c$\ srPBEgws, 
$^{d}$\ Ref. \citenum{fecp2_exp_1971}.}
\end{tabular}
\end{center}
\end{sidewaystable}

\clearpage

%

\end{document}